\documentclass{aa}

\usepackage[T1]{fontenc} 
\usepackage{graphicx} 
\usepackage{booktabs} 
\usepackage{tabularx} 
\usepackage{xltabular} 
\usepackage{multirow} 
\usepackage{xcolor} 
\usepackage{ulem} 
\usepackage{comment} 
\usepackage{csvsimple} 
\usepackage{hyperref} 
\usepackage{natbib} 
\usepackage{orcidlink}
\usepackage{makecell}

\usepackage{dblfloatfix}

\usepackage{float}

\hypersetup{
  colorlinks   = true, 
  urlcolor     = blue, 
  linkcolor    = blue, 
  citecolor   = blue 
}

\newcommand{{\Teff}}{{T_\mathrm{eff}}}
\newcommand{{\logg}}{{\log g}}
\newcommand{{\feh}}{{$[\mathrm{Fe}/\mathrm{H}]$}}

\begin{document}

   \title{Gaia FGK Benchmark Stars: Selecting Infrared Lines for Abundance Determination. \thanks{Based on observations collected at the European Southern Observatory under ESO programmes 1102.22KH, 109.234G.}}

\author{
  S. Elgueta\,\orcidlink{0000-0001-5642-2569}\inst{1,2}\thanks{Corresponding author: \email{sselgueta@gmail.com}}
  \and
  P. Jofr\'e\,\orcidlink{0000-0002-0722-7406}\inst{1}
  \and
  C. Aguilera-G\'omez\,\orcidlink{0000-0002-9052-382X}\inst{3}
  \and
    D. Slumstrup\,\orcidlink{0000-0003-4538-9518}\inst{4,5}
  \and
  Á. Rojas Arriagada\,\orcidlink{0000-0003-0208-8854}\inst{2,6}
  \and
   U. Heiter, \orcidlink{0000-0001-6825-1066}\inst{7}
   \and
  L. Casamiquela, \orcidlink{0000-0001-5238-8674}\inst{8}
    \and
     M. Zoccali, \orcidlink{0000-0002-5829-2267}\inst{3}
 \and
  C. Worley, \orcidlink{0000-0001-9310-2898}\inst{9}
   \and
  C. Soubiran, \orcidlink{0000-0003-3304-8134}\inst{10}
}

\institute{
  Instituto de Estudios Astrofísicos, Facultad de Ingeniería y Ciencias, Universidad Diego Portales, Av. Ejército Libertador 441, Santiago, Chile
  \and
  Departamento de Física, Universidad de Santiago de Chile, Av. Victor Jara 3659, Santiago, Chile
  \and
  Instituto de Astrofísica, Pontificia Universidad Católica de Chile, Av. Vicuña Mackenna 4860, 782-0436 Macul, Santiago, Chile
\and
  Gran Telescopio CANARIAS: Bre\~na Baja - La Palma, Santa Cruz de Tenerife, Spain  
  \and{Instituto de Astrofísica de Canarias, E-38205 La Laguna, Tenerife, Spain}
    \and
  Center for Interdisciplinary Research in Astrophysics and Space Exploration (CIRAS), Universidad de Santiago de Chile, Santiago, Chile
  \and
Observational Astrophysics, Department of Physics and Astronomy, Uppsala University, Box 516, SE-751 20 Uppsala, Sweden
\and
  LIRA, Observatoire de Paris, Université PSL, Sorbonne Université, Université Paris Cité, CY Cergy Paris Université, CNRS,
 92190 Meudon, France
\and
    School of Physical and Chemical Sciences– Te Kura Mat¯ u, University of Canterbury, Private Bag 4800, Christchurch 8140, New Zealand
\and
Laboratoire d’Astrophysique de Bordeaux, Univ. Bordeaux, CNRS, B18N, allée Geoffroy Saint-Hilaire, 33615 Pessac, France
}



\titlerunning{Selecting IR Lines for Abundance Determination}
\authorrunning{Elgueta et al.}
\date{Received <date> / Accepted <date>}

 
  \abstract
{The advent of new and more powerful infrared spectrographs has significantly motivated the advancement of the study of atomic and molecular line lists and stellar atmosphere models. While optical abundance determinations rely on extensively validated line lists and modeling frameworks, infrared measurements still face larger uncertainties, largely driven by the choice of atmospheric models and the quality of the available atomic data. In this work, we aim to deliver a homogeneous and reproducible set of atomic absorption lines in the Y, J, and H bands (9800–18000\,\AA), based exclusively on laboratory atomic data. We analyse CRIRES spectra of six Gaia FGK Benchmark Stars spanning a wide range in effective temperature, surface gravity, and chemical composition. Synthetic spectra are computed using the benchmark stellar parameters, and each transition is evaluated independently in every star through a quantitative sequence that examines line depth, saturation, blending (purity), and the agreement between observed and synthetic line profiles. We identify a set of robust atomic transitions in these bands that remain consistent across the full range of stellar parameters represented in our sample. Lines of $\alpha$-elements such as Mg\,\textsc{i}, Si\,\textsc{i}, and Ca\,\textsc{i}, together with several Fe\,\textsc{i} transitions, satisfy all robustness criteria. Among the neutron-capture species explored, only Sr\,\textsc{ii} provides lines that consistently meet our requirements. Beyond the specific list of accepted transitions, this study demonstrates that a fully quantitative, multi-criteria framework provides a transparent and reproducible foundation for near-infrared line validation as laboratory data, stellar atmosphere models, and instrumentation continue to improve.

}


   \maketitle
   \nolinenumbers

\section{Introduction}

In the current astronomical landscape, we consistently witness new findings in new, distant, and exotic corners of the Universe. This progress has been recently significant, thanks to space-based facilities, in particular, by the infrared (IR) data provided by the James Webb Space Telescope (JWST) \citep{Gardner-2023}.

The JWST has delivered astonishingly deep insights and empirical evidence across a wide range of science programmes, highlighting how essential infrared observations have become for modern astrophysics. However, this progress also underscores that the infrared still lags behind the optical in several aspects, e.g., the quality and completeness of the existing atomic and molecular data, high-resolution spectral libraries, and the techniques required to mitigate the atmospheric effects in ground-based observations. In parallel, developments in ground-based observations in the near-infrared regime have advanced alongside the space programmes and are equally affected by these limitations. The APOGEE survey \citep{Majewski-2017} has revolutionized Galactic archaeology by enabling chemical cartography of the Milky Way on an unprecedented scale and with remarkable robustness in unexplored regions historically obscured by dust \citep[e.g.,][]{Bovy-2014, Hayden-2015, Queiroz-2020}. Furthermore, MOONS \citep{Cirasuolo-2014} will soon begin operations, delivering a vast volume of data over an extended near-infrared (NIR) coverage, not only for stellar physics but also for galaxy evolution studies. This underscores the need for strongly grounded IR information, consistent enough to complement the optical counterparts in facing the new challenges that JWST and MOONS will pose.

Infrared spectroscopic observations offer a significant advantage over optical observations when studying heavily obscured by interstellar extinction \citep{Ryde-2007}. The current facilities and instruments dedicated to mid- to high-resolution infrared spectroscopy have led to the detection of both strong and weak absorption lines. For example, lines of neutral and ionized iron were compiled and studied for a sample of FGK stars observed with the CARMENES spectrograph by \citet{Marfil-2020}, and metal-poor stars were studied in the infrared with the IGRINS spectrograph by \citet{Afsar-2016}. The H band exploration and methodologies developed by the APOGEE survey have allowed astronomers to deepen the understanding of this regime beyond traditional optical constraints. Additionally, the relatively understudied NIR regime, particularly the \textit{Y} and \textit{J} bands hold the potential to yield significant insights into the chemical evolution of the Galaxy due to the presence of a robust collection of n-capture elements that are still not well characterized.

For abundance studies, these broader IR limitations manifest in more practical issues. The lack of complete and homogeneous spectral libraries across the full IR range, together with inconsistencies in basic atomic data such as line positions and intensities \citep{Ryde-2019}, directly affects the identification of reliable absorption features. In addition, abundances derived from different IR bands for the same star can show non-negligible discrepancies. Altogether, this reinforces the need for a well-defined set of lines and consistently constrained stellar parameters when working across the near-IR.

The consistency of these parameters is best evaluated by studying specific stellar archetypes with well-established physical properties. These archetypes correspond to well-known and widely studied samples of stars. In this work, we will explore the IR extension of the Gaia FGK Benchmark Stars (GBS), which represent a thoughtfully curated collection of reference stars crucial for calibrating and validating spectroscopic techniques. These stars cover a wide range of spectral types, including F, G, K, and M at varying luminosities and metallicities \cite{Jofre-2014, Heiter-2015, Hawkins-2016, Jofre-2019, Soubiran-2024, Casamiquela-2025}. They have comprehensive observational data that allow independent determination of their effective temperature and surface gravity with 1-2$\%$ precision. The determination of $\Teff$ and $\logg$ relies on fundamental relationships involving observable quantities such as angular diameters from interferometry, bolometric fluxes, parallaxes, and mass based on theoretical models.

Using the high resolution of CRIRES ($R \sim 90000$; \citealt{Kaeufl-2004}), we analyze six spectra from GBS, each representing a different spectral type as defined in \citet{Jofre-2015}. Our primary goal is to 
identify a global set of reliable absorption lines that provide consistent abundance measurements across a wide range of stellar parameters in the \textit{Y}, \textit{J}, and \textit{H} bands. To achieve this, we provide line recommendations for each spectral type individually. All selections are based on a series of quantitative criteria that will be defined in the following sections, using the Vienna Atomic Line Database (VALD3; \citealt{Ryabchikova-2015}) as our atomic reference. 

Recent work by \cite{Senturk-2024} compiled a detailed NIR (\textit{YJHK}) line list using the Kitt Peak National Observatory (KPNO) spectra of the Sun and a solar analog as references. While valid for solar analogs, their calibration does not account for the severe line blending and atmospheric changes present in cooler or evolved stars. By validating our list against the diverse Gaia FGK Benchmark standards, we overcome this limitation, ensuring reliability across the HR diagram.

Extending such validity is particularly relevant in the infrared, where the identification of suitable lines for accurate abundance determination is significantly more challenging than in the optical regime, for several reasons. First, the number of measured atomic (in laboratory) transitions is far more limited in the IR, especially regarding accurate $\log gf$ values and collisional broadening parameters \citep{Ryde-2009, Ryde-2010, Shetrone-2015}. Second, molecular contamination and the overall density of spectral absorption features increase in cooler stars, causing severe line blending and greater uncertainty in continuum identification. 
Third, benchmark datasets in the IR remain comparatively sparse relative to those in the optical, which complicates efforts to validate or calibrate spectral lines across a range of stellar types \citep{Thorsbro-2020}. As a result, the effectiveness of traditional line-by-line differential abundance methods becomes increasingly constrained in this regime, unless the selected lines are demonstrably robust in terms of depth, blending, and model sensitivity.
Consequently, focusing on lines that remain stable under standard assumptions is essential to ensure the physical consistency and reproducibility of our abundance analysis without relying on object-dependent fine-tuning.

Because we aim to work with a line list anchored to laboratory atomic data, we do not use empirically calibrated line lists, such as APOGEE's as a primary reference. Their $\log gf $ values are adjusted to match the observations, making them overall effective for large-scale survey pipelines but not necessarily suitable for the analysis with a different pipeline or instrument than APOGEE itself. Developing an independent list ensures that the selection criteria applied in this work rest on a uniform, reproducible foundation.

This paper is structured to follow the complete selection process that each spectral line undergoes before being accepted into the final robust line list. Rather than presenting a purely static methodology, we illustrate the method by tracing the journey of two absorption lines: Si\,\textsc{i} (10786.849 \AA) and Sr\,\textsc{ii} (10036.653 \AA) in the \textit{Y} band across different spectral types. These lines were selected to illustrate the method's performance on two distinct cases: a standard $\alpha$-element widely used in abundance studies, and a key $n-$capture element. At each step, using depth evaluation, saturation checking, purity assessment, and fit quality, we show how these spectral features are progressively filtered, accepted, flagged as uncertain, or ultimately rejected. Section~\ref{sec:obs} describes the observational data and synthetic spectra. Section~\ref{sec: Method} details the line selection criteria, including the threshold definitions and quality flags. Section~\ref{sec:robust} presents the results of applying these filters into a Robust set of lines for each of the spectral types. Finally, Section~\ref{sec:discussion} provides a summary and future prospects.

\section{Observations and Data Preparation}
\label{sec:obs}

We used the CRIRES+ instrument with ESO's VLT at Paranal Observatory, Chile. This upgraded version of the CRIRES instrument (CRyogenic InfraRed Echelle Spectrograph) is a cross-dispersed echelle spectrograph operating in the mid- and near-infrared (\textit{YJHKLM} bands) at a nominal spectral resolution of $\sim$100000 with the 0.2 arcsecond slit \citep{Dorn-2023}.
We analyzed spectra for a representative sample of six GBS, including a solar spectrum obtained from observations of Vesta. Our sample spans a wide range in stellar parameter space ($\Teff$, $\logg$, and \feh), ensuring broad coverage for this study. These observations were conducted under the program ID:102.22KH and 109.234G. The stellar parameters for the selected targets are listed in Table~\ref{tab:gbs_all_vertical}, and were derived from angular diameters, bolometric fluxes, and evolutionary models based on Gaia data \citep{Soubiran-2024}. For further details on the GBS selection and methodology, we refer the reader to \citet{Soubiran-2024} and references therein.

\begin{figure*}
    \centering
    \includegraphics[width=\linewidth]{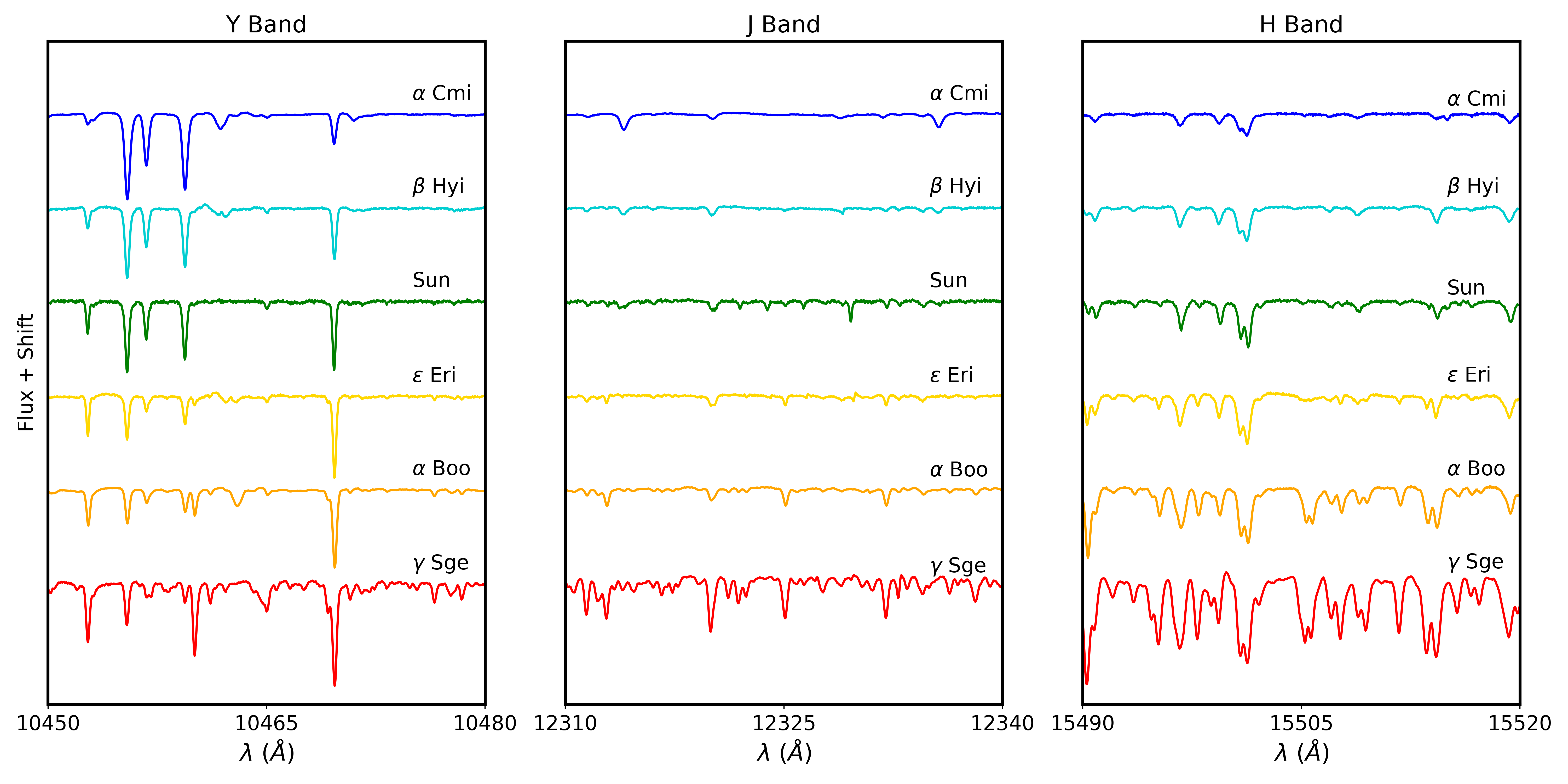}
    \caption{Spectral cuts for each of the three bands are shown for the representative stars, after telluric correction, RV correction, and normalization. The stars are sorted by temperature, making clear how the line profiles evolve from warmer to cooler GBS.}
    \label{fig:3bands-plot}
\end{figure*}

 In this analysis, we focus on three working regions in the \textit{Y} band: $[9800 - 10800]$ $\text{\AA}$, the \textit{J} band: $[11800 - 13200]$ $\text{\AA}$, and the \textit{H} band: $[15000-17500]$ $\text{\AA}$. In Figure~\ref{fig:3bands-plot}, we plot a small range of the spectrum of each band for all our sample, the spectra shown are telluric corrected with \texttt{MOLECFIT}, normalized and radial velocity (RV) corrected. These infrared regions, particularly the \textit{Y} and \textit{J} bands \citep{Sharon-2010, Matsunaga-2020}, remain relatively unexplored in a homogeneous and systematic way, yet they contain a broad variety of atomic transitions of astrophysical interest. This includes several $n$-capture species (e.g. Sr\,\textsc{ii}, Y\,\textsc{i}, Y\,\textsc{ii}, La\,\textsc{ii}, Ce\,\textsc{ii}), as well as elements such as Al\,\textsc{i}, P\,\textsc{i}, S\,\textsc{i}, and Zr\,\textsc{i}. As shown throughout this work, only a subset of these transitions satisfies all the quantitative selection criteria adopted here; those lines are the ones that can be used with confidence for abundance determinations across different stellar types and wavelength bands.

 \begin{table}[t]
\renewcommand{\arraystretch}{1.2} 
\centering
\footnotesize
\caption{Stellar parameters for the  GBS used in this study from \cite{Soubiran-2024}, grouped by evolutionary stage.}
\label{tab:gbs_all_vertical}
\begin{tabularx}{\linewidth}{l *{3}{>{\centering\arraybackslash}X}}
\toprule
\multicolumn{4}{c}{\textbf{\textit{Dwarfs}}} \\
\midrule
\textbf{Parameter} &
\shortstack{$\alpha$ CMi \\ {\scriptsize F Dwarf}} &
\shortstack{Vesta/Sun \\ {\scriptsize G Dwarf}} &
\shortstack{$\epsilon$ Eri \\ {\scriptsize K Dwarf}} \\
\midrule
$T_\mathrm{eff}$ (K)            & 6582  & 5778   & 5130 \\
$\sigma~T_\mathrm{eff}$         & 5.0   & 1.0    & 30.0 \\
$\log g$ (dex)                  & 3.98  & 4.44   & 4.63 \\
$\sigma~\log g$                 & 0.02  & 0.0002    & 0.01 \\
{[Fe/H]} (dex)                  & -0.02 & 0.02   & -0.09 \\
$\sigma~[\mathrm{Fe}/\mathrm{H}]$ & 0.02  & 0.01    & 0.01 \\
OBS Date   & 2022-04-06       &   2022-11-22     &   2021-11-16   \\
UT   &   02:11 & 00:23 & 01:30  \\
S/N$_Y$                         & 659.7 & 220.45 & 263.18 \\
S/N$_J$                         & 447.34& 155.03 & 263.30 \\
S/N$_H$                         & 359.47& 152.26 & 192.51 \\
EXPTIME (s)      &   12.9    &   362.9     &  51.5 \\

\addlinespace
\midrule
\multicolumn{4}{c}{\textbf{\textit{Giants and Subgiants}}} \\
\midrule
\textbf{Parameter} &
\shortstack{$\beta$ Hyi \\ {\scriptsize FGK Subgiant}} &
\shortstack{$\alpha$ Boo \\ {\scriptsize FGK Giant}} &
\shortstack{$\gamma$ Sge \\ {\scriptsize M Giant}} \\
\midrule
$T_\mathrm{eff}$ (K)            & 5917  & 4277   & 3904 \\
$\sigma~T_\mathrm{eff}$         & 25.0  & 23.0   & 30.0 \\
$\log g$ (dex)                  & 3.97  & 1.58   & 1.06 \\
$\sigma~\log g$                 & 0.04  & 0.07   & 0.04 \\
{[Fe/H]} (dex)                  & -0.12 & -0.55  & -0.26 \\
$\sigma~[\mathrm{Fe}/\mathrm{H}]$ & 0.02  & 0.01   & 0.06 \\
OBS Date      & 2021-10-28     &   2021-10-28       &  2022-05-16      \\
UT   &  03:05 & 02:11 &  07:50 \\
S/N$_Y$                         & 308.68& 470.4  & 103.87 \\
S/N$_J$                         & 260.95& 547.34 & 103.7 \\
S/N$_H$                         & 248.80& 381.67 & 37.84 \\
EXPTIME (s)   &   51.5    &  2.9      &   35.5   \\
\bottomrule
\end{tabularx}
\end{table}

Although space-based facilities such as the JWST provide access to wavelengths overlapping the classical \textit{L} and \textit{M} bands, their spectroscopic capabilities are not optimized for high-resolution stellar abundance work in this regime. In particular, NIRSpec operates over the range $\sim$0.6-5.3\,$\mu$m, partially covering the \textit{L} band, but at spectral resolutions that are generally lower than those of ground-based high-resolution spectrographs. At longer wavelengths, MIRI covers $\sim$4.9 - 28.8\,$\mu$m, including the \textit{M} band, but is primarily designed for low- to medium-resolution spectroscopy in the mid-infrared.

As a result, current high-resolution stellar abundance studies and large spectroscopic surveys remain primarily focused on the near-infrared \textit{Y}, \textit{J}, and \textit{H} bands, where both instrumental performance and legacy value are maximized.

For the reduction and wavelength calibration of our data set, we used the CRIRES+ pipeline in the ESO-Reflex environment \citep{Freudling-2013}. Furthermore, in IR spectroscopy, we need to account for an extra (and crucial) step, the telluric correction. Infrared spectral analysis is particularly vulnerable to telluric contamination, which arises from the Earth's atmosphere. The presence, shape, and intensity of telluric lines depend on factors such as the observatory's location, atmospheric conditions, and airmass \citep{Noll-2012, Smette-2015}. To ensure accurate identification and analysis of stellar features, it is crucial to eliminate telluric contamination from the observed spectra.

Although our observations were carried out under the exceptional conditions of Paranal Observatory in Chile, atmospheric absorption lines remain present and must be carefully understood, modeled, and removed. The challenge lies in the intrinsic time-variable nature of the atmosphere, which makes its behavior difficult to predict and therefore complicates the possibility of making the correction procedure more automatic, if not completely automatic \citep{Ulmer-Moll-2019, Gullikson-2014}.

Various methods have been developed to mitigate this issue. These range from physical modeling based on the known molecules, such as HITRAN \citep{Gordon-2022}, to data-driven or semi-supervised approaches that make use of observational metadata (e.g., airmass, humidity, pressure) recorded in the FITS headers to generate synthetic atmospheric transmission spectra. However, residual uncertainties persist, especially in spectral regions densely populated by strong telluric bands. A widely adopted strategy to improve correction accuracy is to observe a telluric standard star,  typically an A0V star \citep{Sameshima-2018} or a rapidly rotating G Dwarf star \citep{Maiolino-1996, Vacca-2003}, immediately before or after the science target, thereby minimizing the impact of time variability. While effective, this approach is observationally expensive and impractical for large samples. 

Instead, we adopted a semi-automated, yet interactive method based on the \texttt{MOLECFIT} software \citep{Smette-2015}, in its standalone version. In this work, we applied the telluric correction to both full-band extractions and selected CRIRES+ detector chips containing the regions of interest within the \textit{Y}, \textit{J}, and \textit{H} bands.

The \textit{Y} band offers an almost pristine (telluric-free) atmospheric window spanning from 10280 to 10680 [\AA], which requires minimal telluric correction. After normalization, this \textit{Y}-band atmospheric window allows for the direct identification and confirmation of spectral lines, especially those associated with heavy metallic lines and n-capture elements, as documented in previous studies \citep{Elgueta-2024, Matsunaga-2020}.
It is, however, not replicated in the \textit{J}, and \textit{H} bands. In fact, the most dramatic contamination by telluric features can be observed in the H band (APOGEE's and MOON's regime) for which we opt to skip some regions given the difficulty encountered in the removal of the telluric features, e.g. nonphysical quantities for the molecules' column densities, and re-normalization of the continuum.

The chosen spectral intervals in the \textit{Y}, \textit{J}, and \textit{H} bands are particularly suitable, as they contain portions that can be effectively corrected. Nevertheless, some spectral windows remain resistant to accurate correction without a dedicated telluric standard. To assess the success of our correction, we measured the residual depth of the telluric features after correction. For CRIRES data, this typically falls below 5$\%$, giving us confidence that most of the absorption lines included in our analysis lie outside significantly contaminated regions or are located in effectively corrected intervals.

Once the telluric absorption lines were removed from the spectra with \texttt{MOLECFIT}, the spectra are normalized using the \texttt{iSpec} software \citep{Blanco-Cuaresma-2014}\footnote{\url{https://www.blancocuaresma.com/s/iSpec}} by fitting a spline with order (n) equal to 3 to each spectral band. A direct division of the spectrum by a smoothed version of itself was avoided, since it can bias the continuum placement in regions with high line density. This consideration is particularly important in the NIR, where molecular absorption, residual telluric structure, and the instrumental response can change the pseudo–continuum, potentially leading to inaccurate line depths or artifacts. Thus, the spline method provides a more stable and reproducible continuum placement across different spectral types.

The RV correction was obtained by measuring the Doppler shift via cross-correlation, using the \texttt{crosscorrRV} routine\footnote{\url{https://pyastronomy.readthedocs.io/en/latest/pyaslDoc/aslDoc/crosscorr.html}}. 
We note that \texttt{iSpec} implements RV determination based on the same cross-correlation principle; however, we did not use it because its default template set and configuration do not provide homogeneous coverage for our working bands and spectral types. 
Instead, we performed the cross-correlation directly with \texttt{crosscorrRV}, adopting reference templates from the \texttt{iSpec} library: an observed solar spectrum for the Y and J bands, and a synthetic model ($T_{\mathrm{eff}} = 5777$\,K, $\log g = 4.44$, [M/H]$=0.0$) for the H band. Additional synthetic templates were calculated for each of the remaining five spectral types in our sample. After these later stages in correction, the resulting spectra are prepared for the analysis presented in this work, which will be described in further sections.

\section{Line Selection}
\label{sec: Method}

It is worth noting that excellent line compilations already exist in other regimes, such as the optical work by \citet{Heiter-2021}, as well as in the IR with the APOGEE list \citep{Shetrone-2015}, and other studies from \cite{Afsar-2016, Marfil-2020, Senturk-2024}. 
The APOGEE linelist, in particular, has been crucial for large-scale infrared surveys, but it is largely based on astrophysical fine-tuning to reproduce observed spectra in a survey framework. 
For this reason, we do not adopt APOGEE as the input linelist in this work. Instead, our line selection is anchored to laboratory atomic data and is validated through the uniform, quantitative criteria described below, exploiting the very high resolution of CRIRES in the comparatively less explored \textit{Y}, \textit{J} and \textit{H} bands.

\begin{figure}[h]
    \centering
    \includegraphics[width=\linewidth]{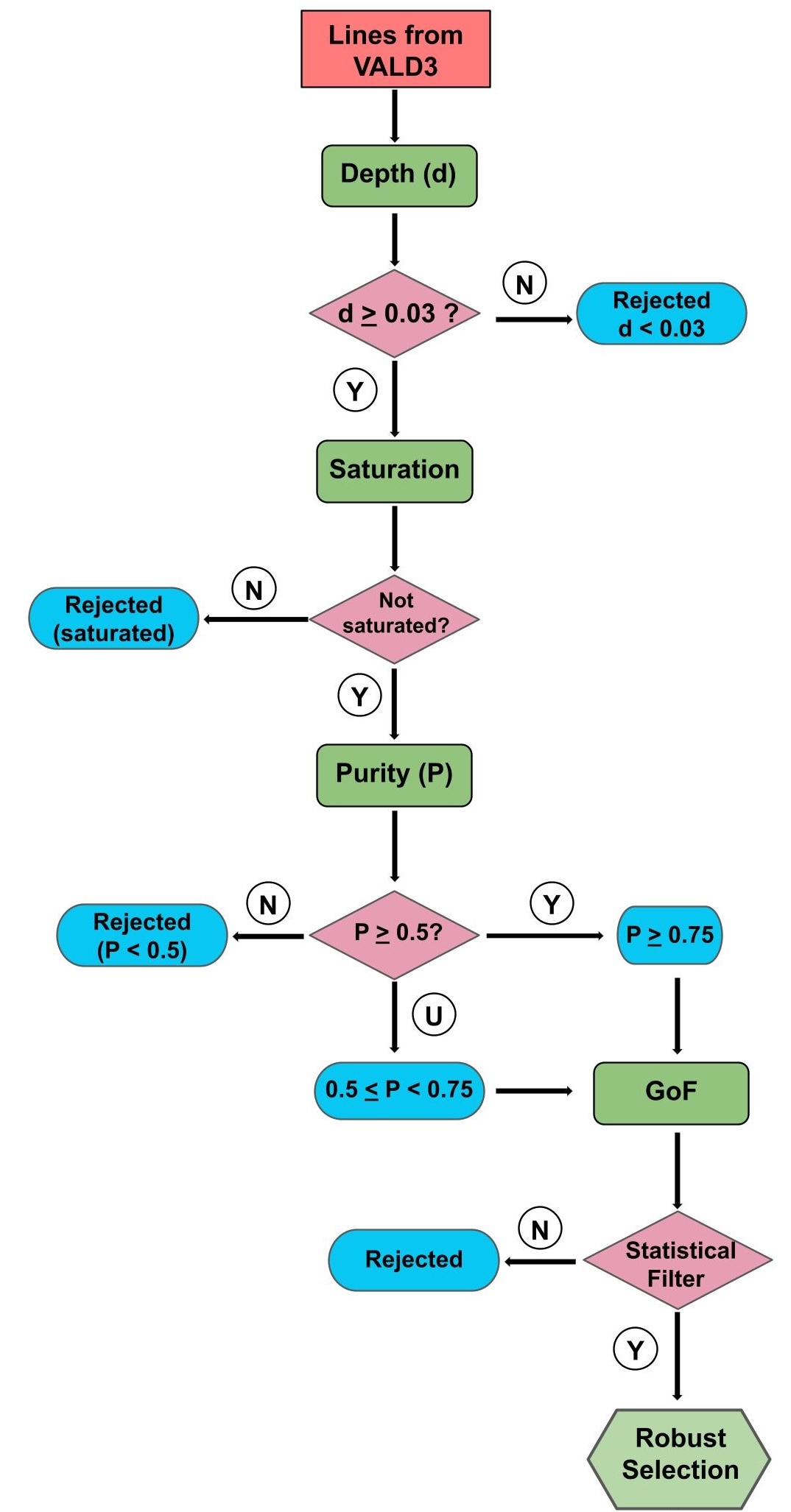}
    \caption{Methodology flow chart where the flags for each process are defined.}
    \label{fig:ir_flow}
\end{figure}

To ensure homogeneity, we compare the sample in Table~\ref{tab:gbs_all_vertical} with synthetic calculations based on laboratory atomic data from VALD, rather than relying on solar or astrophysically calibrated line lists. The goal is not to recalibrate atomic data or stellar parameters \citep[e.g.,][]{Elgueta-2024}, but to provide a robust set of lines that remain physically stable as abundance indicators and reproducible across different spectral qualities without requiring further adjustments, following the methodology described in Figure~\ref{fig:ir_flow}.

Our analysis follows five consecutive steps, which will be described in the following subsections.
(i) We synthesize a synthetic spectrum with the stellar parameters ($\Teff$, $\logg$ and \feh) from \cite{Soubiran-2024}, broadening parameters from \cite{Jofre-2014, Heiter-2015}, and the abundance measurements in optical from \cite{Casamiquela-2025}. (ii) We measure the central depth of each candidate line, keeping only those with depth values $\geq 0.03$. 
(iii) The lines passing this threshold are then examined for signs of saturation, combining the curve of growth with an inspection of their widths through spectral synthesis. 
(iv) Lines that are not saturated proceed to a purity assessment, where we evaluate the degree to which each feature is affected by neighbouring blends (see Section~\ref{sec:purity} for the formal definition)
(v) Features that pass this blending check are then tested for their goodness of fit (GoF) to the observed spectra, following the statistical criteria described in Section~\ref{sec:gof}.
This sequence (see Figure~\ref{fig:ir_flow}) delivers a final set of lines that can be considered suitable for abundance determinations in the \textit{Y}, \textit{J} and \textit{H} bands, whose analysis will be explained in more detail in the following sections.

\subsection{Synthesis}

The synthesis was carried out using a python wrapper for MOOG \cite{Sneden-2012} developed by Jian, M. \footnote{\href{https://github.com/MingjieJian/pymoog}{Pymoog's github}} employing the MARCS Spherical with Local Thermodynamic Equilibrium (LTE) models \citep{Gustafsson-2008}. 
Our VALD linelist was extracted in 2015-01-21, version 968.

The use of LTE or NLTE in 1D and/or 3D models responds uniquely to the fact that the departure coefficients needed for each transition are widely calculated for optical regimes but not comprehensively for the infrared regime, yet some efforts to correct abundances in APOGEE spectra have been made (see, e.g., \citealt{Osorio-2020}). 

\begin{figure}[t]
    \centering
    \includegraphics[scale=0.5]{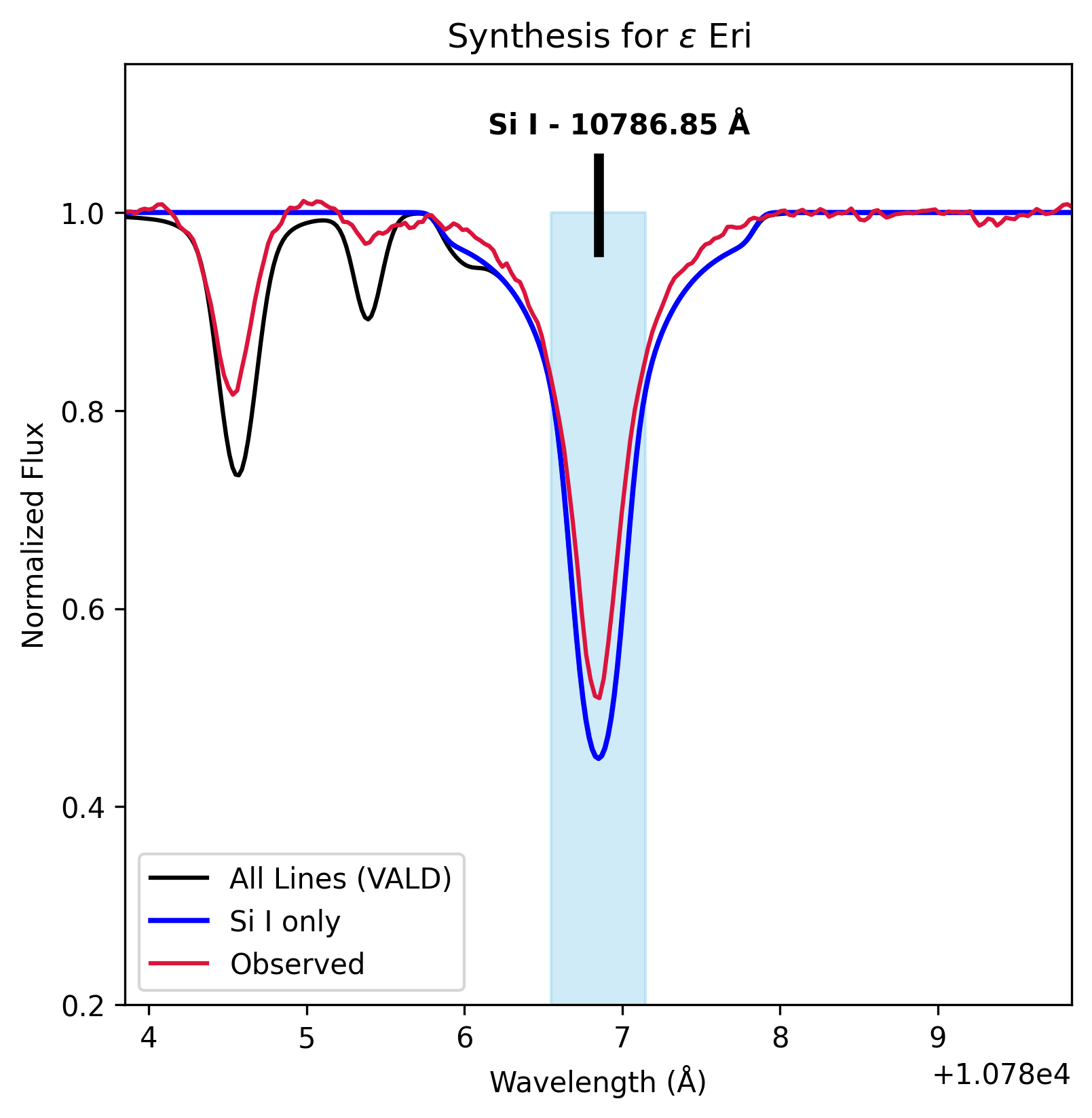}
    \caption{Synthesis of the  Si\,\textsc{i} absorption line for $\epsilon$ Eri (K Dwarf) in the Y band. The line in crimson indicates the synthesis for  Si\,\textsc{i} only. }
    \label{fig:synth_si1_epsilon}
\end{figure}

The spectral synthesis is run in two configurations: first, a full synthesis including all elements present in the linelist within the specified wavelength range (hereafter referred to as "All"); and second, a synthesis including only the element of interest ("element only"). 

It is worth noting that for $[X/H]$ values, where individual optical abundances are unavailable (e.g., S\,\textsc{i}), we follow a practical approach by scaling $[X/H]$ to the star's global metallicity $[M/H]$ from \cite{Casamiquela-2025}. 
In particular, for the case of  \ion{S}{I}, treating it as an \(\alpha\)-element is also supported by observational studies showing that its abundance generally follows the \(\alpha\)-pattern in disk and bulge stars \citep{Caffau-2005, Nissen-2007, Spite-2011}. Therefore, the resulting purity estimates ($P\equiv \mathrm{EW}_1/\mathrm{EW}_2$) are expected to remain robust, with the main caveats applying to the weakest features and to lines approaching saturation \citep[see, e.g.,][]{Gray-2005}.

Molecular line lists were not included in the synthesis; however, at the temperature of our coolest target ($\gamma$ Sge, $T_{\mathrm{eff}} = 3904$ K) molecular absorption becomes significant and can complicate the analysis. We nevertheless chose to include $\gamma$ Sge in our sample, despite these challenges, as it provides a valuable reference point for the applicability of our atomic line-selection criteria toward the coolest GBS stars. An example of an atomic synthesis, highlighting the transition under study, is shown in Figure~\ref{fig:synth_si1_epsilon}. In this case, the observed feature appears slightly shallower than the one predicted by the synthetic calculation, pointing to non-negligible discrepancies in the atomic data or in the modeling of the Si\,\textsc{i} transition itself. Comparing the two synthetic spectra with the observations provides a qualitative check on how reliable a line is, which, in tandem, motivated the creation of solely quantitative criteria introduced in the next sections.

\subsection{Flagging}
\label{sec:flagging}

We propose a flagging system similar to the one in \cite{Heiter-2021}, but divided into four stages (or flags). 
For each line of interest, we assign flags for: Depth (\(\texttt{d}\)), Saturation, Purity (\texttt{P}), and a final synthesis/fit quality step, understood as the goodness of fit (GoF). 

At every stage, a line may receive one of three possible labels: \texttt{Y} (``Yes'', the criterion is satisfied and the line proceeds), 
\texttt{N} (``No'', the criterion fails and the line is rejected), and 
\texttt{U} (``Undecided''), meaning that the line does not fully satisfy the criterion but remains viable and is therefore passed to the next step 
of the decision tree). The \texttt{U} flag can only be assigned in the purity stage; 
all other tests are strictly binary (\texttt{Y} or \texttt{N}).

The purity flag plays a role similar to the synflag defined in \cite{Heiter-2021}, where the impact of blends is quantified, whereas the synthesis/fit quality flag assesses the agreement between the observed and synthetic profiles using statistical metrics (e.g. $\chi^2$ ) and is used as a consistency check in our decision tree.

In principle, the combination of the binary and categorical criteria (depth, saturation, purity, and statistical filter) yields $2 \times 2 \times 3 \times 2 = 24$ theoretical configurations. In practice, however, the criteria are applied sequentially in a decision-tree (see Figure~\ref{fig:ir_flow}, which excludes many of these combinations). Lines rejected at an early stage (e.g. insufficient depth, or saturated) do not proceed to the following steps, and only those that pass continue to subsequent evaluations. Consequently, the number of effective outcomes is reduced to seven, corresponding to the distinct terminal branches of the decision tree.

 Among these, we prioritize combinations that lead to favorable outcomes for our analysis, specifically those with final decisions labeled as \texttt{YYYY} or \texttt{YYUY}. These latter represent lines with sufficient depth, not saturated, acceptable purity, or blending conditions, and a robust fit between the synthetic and observed spectra given the fixed stellar parameters and abundances. 
The line selection process followed in this work summarized in Figure~\ref{fig:ir_flow} starts with the measurement of depth ($\texttt{d}$) in the synthetic spectra, thus lines with $\texttt{d} < 0.03$ are rejected, as they are considered too shallow to provide reliable measurements. 
In order to decide whether a line is saturated, we rely on the classical definition of the curve of growth: in the linear regime the equivalent width increases proportionally with abundance, while in the saturation regime this growth slows down and eventually flattens, so that the line no longer provides reliable information on the abundance \citep[e.g.][]{Gray-2005, Mihalas-1978, Magain-1984}. To capture this behavior in a simple quantitative way, we measure the local slope of the curve of growth around the reference abundance and the change in the line-core flux. If the slope drops below 0.5, meaning that the line has entered a plateau-like regime, and at the same time the core flux varies by less than 0.5$\%$ when increasing the abundance by 0.2 dex, we flag the line as saturated. These thresholds are empirical but follow directly from the definition that once a line stops responding to changes in abundance, it ceases to be a useful diagnostic of elemental abundances. Non-saturated lines were then evaluated in terms of their purity ($\texttt{P}$), defined as the fractional contribution of the target element to the total line strength. Lines with $\texttt{P} < 0.5$ are rejected. Lines with $0.5 \leq \texttt{P} < 0.75$ (\texttt{U}) and those with $\texttt{P} \geq 0.75$ (\texttt{Y}) are both passed to the statistical filter of the Goodness of Fit (GoF), in which they will be differentially tested for each of the bands to assess how well the synthesis does reproduce the observation.

\subsubsection{Depth}
\label{subsub:depth}

\begin{figure*}
    \centering
    \includegraphics[width=\linewidth]{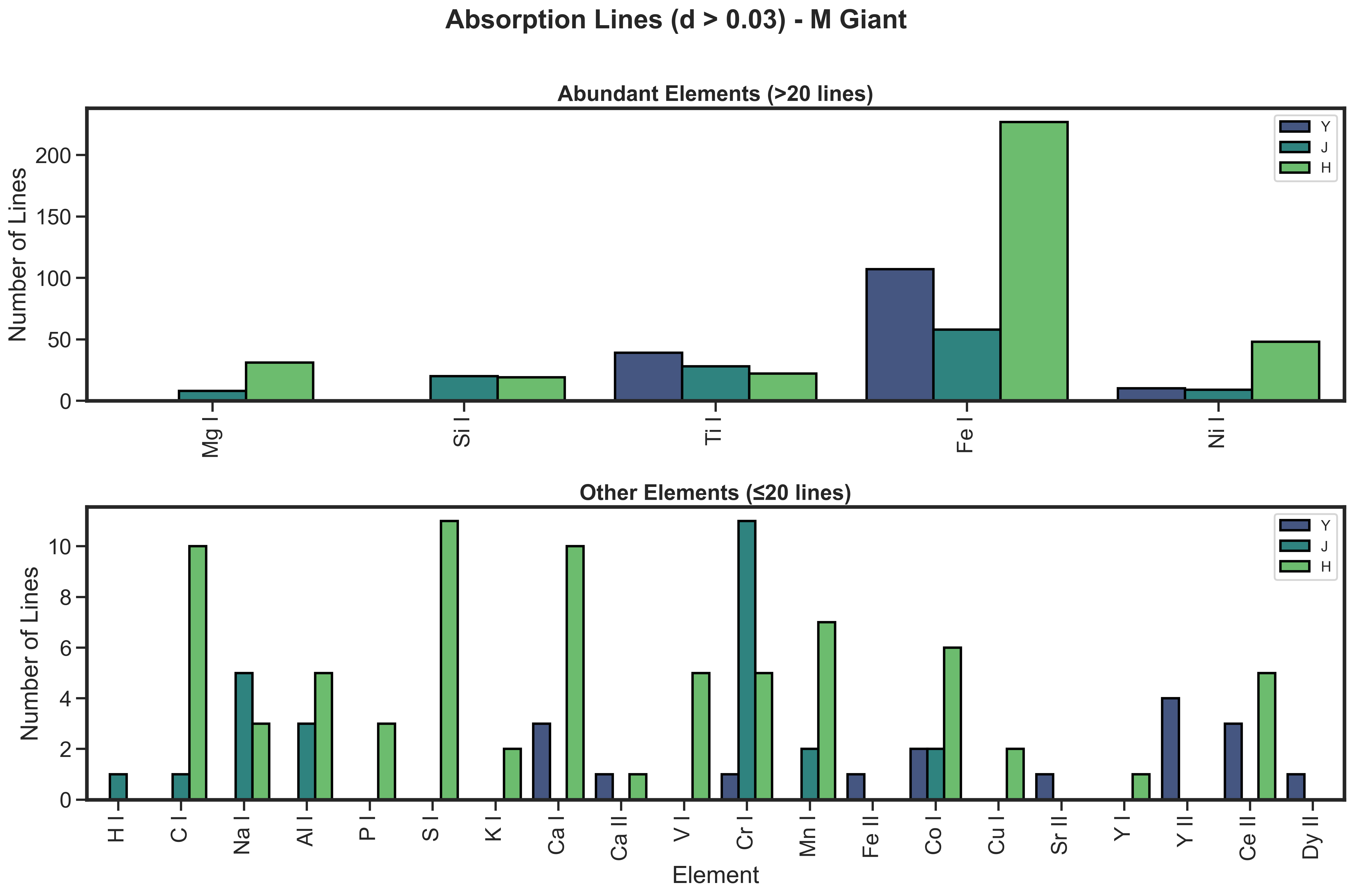}

    \vspace{0.1cm}

    \includegraphics[width=\linewidth]{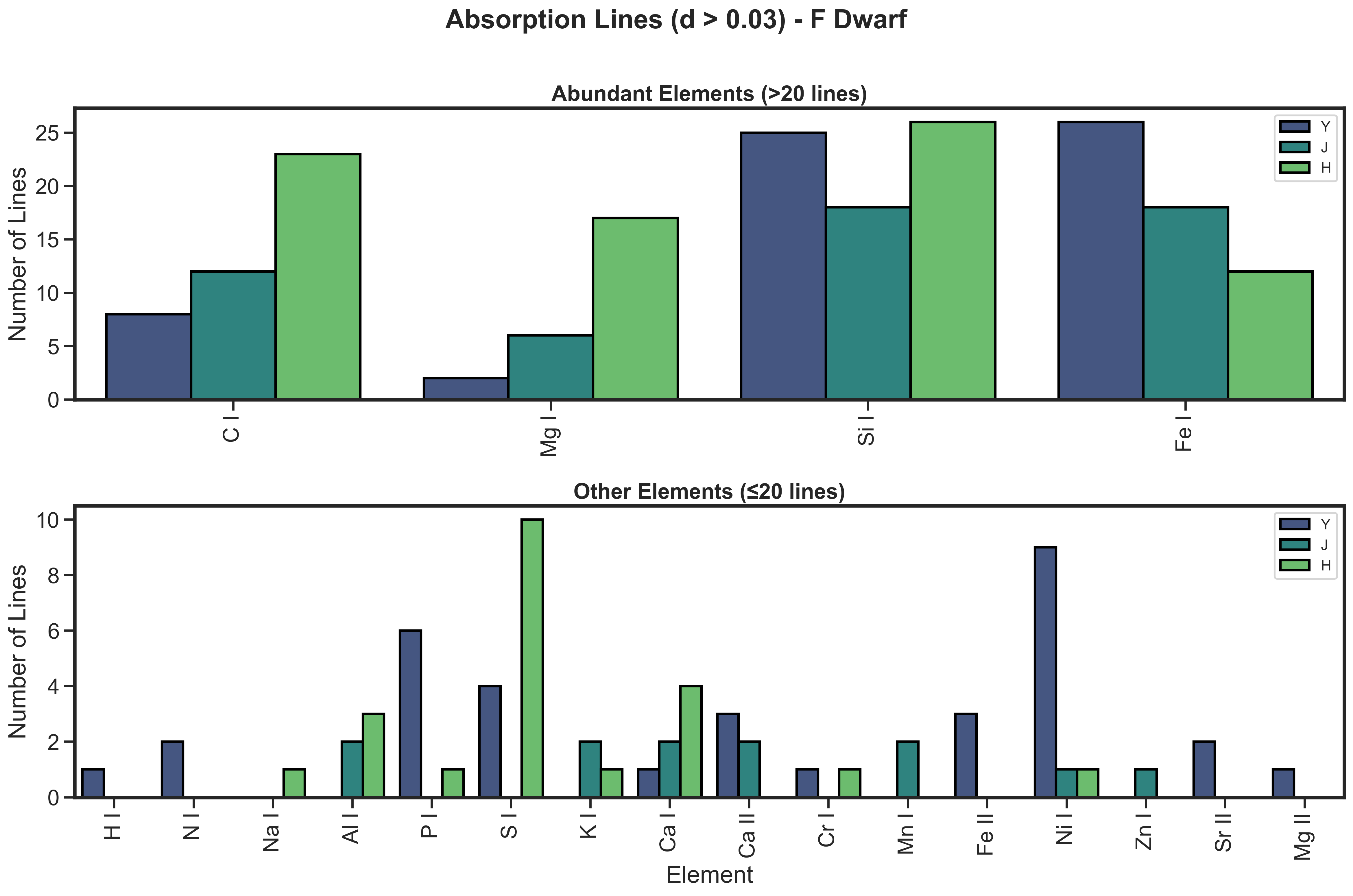}

    \vspace{-0.2cm}

    \caption{Number of absorption lines with $d > 0.03$. For both M Giant and F Dwarf.}
    \label{fig:depth_MGiant_FDwarf}
\end{figure*}

The central line depth (\texttt{d}) of an absorption line is a useful diagnostic of astrophysical conditions such as temperature sensitivity, ionization balance, and pressure broadening \citep{Gray-2005}. A \texttt{d} cut-off of 0.03 ensures our line diagnostics remain both sensitive and robust across spectra with S/N as low as 38 in the \textit{H} band and up to over 650 in the \textit{Y} band (the maximum value within our sample).

For a given $d$, the corresponding $1\sigma$ uncertainty in the depth, $\sigma$, is set by the continuum S/N through Eq.~\ref{eq:errindepth}. For S/N $\sim 100$, Eq.~\ref{eq:errindepth} yields $\sigma \approx 0.01$, so our adopted depth threshold of $d = 0.03$ corresponds to a $\sim 3\sigma$ detection. For most of our spectra, where the S/N typically exceeds 200, this depth threshold implies an even higher detection significance, even in the lower S/N regimes. This depth threshold was also adopted in \citet{Casamiquela-2025}, where a similar S/N significance criterion was used, minimizing the likelihood of spurious detections while preserving the sensitivity to astrophysically meaningful stellar features.

To measure the depth of each line confirmed in the synthetic spectrum, a Gaussian fit was performed over a window of five pixels centered on the position of the theoretical line ($\lambda_c$). The wavelength of the line minimum ($\lambda_0$) was then determined directly from the synthetic spectrum, whose continuum is normalized by definition. 
$d$ was calculated as the difference between the continuum level (unity) and the flux at $\lambda_0$. 

Measurements in the observed spectrum were discarded if the continuum normalization was incorrect (resulting in negative depths), or if the location of $\lambda_0$ deviated significantly from $\lambda_c$.
While the depth is measured in the synthetic spectrum, we also examine the corresponding features in the observed spectra to identify lines whose empirical depths deviate from their theoretical predictions, observed measurements also follow the same $d$ limit of $0.03$. This procedure was carried out using the $ir\_ldr$ python package developed by Jian, M. \footnote{\href{https://github.com/MingjieJian/ir_ldr}{Available at ir\_ldr's github}}. The error in each observed depth measurement is linked to the signal-to-noise ratio (S/N) for the target (obj) as described by equation~\ref{eq:errindepth}:
\begin{equation}
    \sigma = \frac{1}{(S/N)_{\mathrm{obj}}}
    \label{eq:errindepth}
\end{equation}

Columns 3 and 4 in table~\ref{tab:depth_purity_gof_flags} provides an example of how the first flag decision (Y or N) for the depth measurement is determined. For most cases, the observed depth agrees well with the synthetic prediction, provided that the line is captured within the spectral coverage. An exception is seen where no meaningful absorption is detected, the algorithm assigns a default placeholder value (e.g., 1.0 or negative), which is subsequently flagged as non-detected and excluded from the next steps.

It is worth noting that the line depth serves as a preliminary diagnostic of NLTE effects, since stronger lines typically exhibit larger departures from LTE \citep{Lind-2012,Bergemann-2012}. Consequently, many studies in the optical regime have imposed explicit depth thresholds to limit NLTE biases for example, \citet{Mashonkina-2011} for Fe I lines in metal‐poor stars, \citet{Amarsi-2016} for the O\,\textsc{i} triplet, and \citet{Sitnova-2015} for Ti\,\textsc{i} transitions.

Even in the optical, full 3D NLTE corrections can reach up to 0.1 dex for moderate to strong lines, underscoring the limitations of purely depth‐based selection \citep{Bergemann-2017}. In the infrared regime, the situation becomes more complex as the IR lines form deeper in stellar atmospheres and involve uncertain collisional data, which, in tandem, can enhance NLTE deviations \citep{Zhang-2016,Amarsi-2020}. Recent NLTE studies of Si\,\textsc{i} and Ca\,\textsc{ii} triplet lines in the IR have reported abundance corrections of up to 0.2 dex in FGK stars, demonstrating that simple depth cuts may be insufficient \citep{Zhang-2016, Amarsi-2020}. Hence, while depth thresholds remain as useful initial filters, robust abundance analyses in the IR regime require comprehensive thermodynamic‐equilibrium calculations and, ideally, full 3D NLTE modeling, which is not yet available for all the elements of interest in this study.

Nevertheless, in this work, we proceed with our analysis, acknowledging that different physical factors may influence the behavior of lines with depths $d \geq 0.03$. Particular attention will be given to absorption lines that display large values in depth, such as those of Sr\,\textsc{ii} and Si\,\textsc{i} found in the Y and J bands. We exclude lines for which the absorption minimum is ambiguous or significantly offset from the expected central wavelength, $\lambda_{\rm c}$, in the synthetic spectra of the representative stars.

Figure~\ref{fig:depth_MGiant_FDwarf} provides a visual summary of the number of lines that meet the depth threshold ($d > 0.03$) in the \textit{Y}, \textit{J}, and \textit{H} bands for the M Giant representative ($\gamma$ Sge). The upper and lower panels display, respectively, the elements with more than 20 qualifying lines and those with 20 or fewer, highlighting both the dominance of Fe\,\textsc{i} and the diversity of other detectable species across the infrared spectrum. In a more global evaluation, when we compare two different spectral types, such as the M Giant and F Dwarf, we see how the M Giant shows a markedly richer set of neutral lines due to its lower ionization fraction, whereas the F Dwarf exhibits fewer detectable transitions. This systematic variation with $\Teff$ emphasizes the need to evaluate line presence across different spectral types and bands, when building the IR line list.

\subsubsection{Saturation}
\label{subsec:saturation}

In our line selection procedure, we define saturation as the condition where the line response to variations in chemical abundance (\texttt{X}) is negligible (or null) both in the curve of growth and in the line core. For each transition, synthetic spectra at multiple abundances are generated, and saturation is flagged when the local COG slope at the reference abundance \citep{Casamiquela-2025} falls below ${\rm d}\log(W/\lambda)/{\rm d}X < 0.5$, where \texttt{W} is the equivalent width and $\lambda$ is the wavelength. Additionally, the change in core flux is $\Delta F_{\rm core} < 0.005$. This limit is associated with the sensitivity limit that a spectrum with S/N $\sim$ 200 can have. Since our sample has a wide range of S/N values ($\sim$38-$\sim$660), this criterion guarantees that the selected lines have an intrinsic response to abundance variation that is distinguishable from the noise floor.

To quantify saturation and place each transition on the COG, we synthesize a narrow window of $\pm5$\,\AA\ around the target wavelength with \texttt{pymoog} at $R=90000$, adopting the star’s
$T_{\rm eff}$, $\log g$ from Table~\ref{tab:gbs_all_vertical}.  This strategy is consistent with standard practices in widely used spectroscopic frameworks, which typically rely on scaled-solar or \(\alpha\)-enhanced abundance patterns, and fit global parameters such as \([{\rm M/H}]\) and \([\alpha/{\rm M}]\) either before or alongside the determination of individual elemental abundances \citep{Blanco-Cuaresma-2014, Gustafsson-2008, Garcia-2016}. 

For each line of element ID $=$ \texttt{element\_id}, we compute a grid of synthetic spectra at abundance offsets
$\Delta X\in\{-2,\ldots,2\}$ [dex], using the same $\pm5$,\AA\ spectral window, and we assess the line saturation relevant for the COG by inspecting the synthetic line profile within a narrower interval within $\pm 1$\,\AA\ window around $\lambda_c$.

\begin{figure}
 \includegraphics[width=\linewidth]{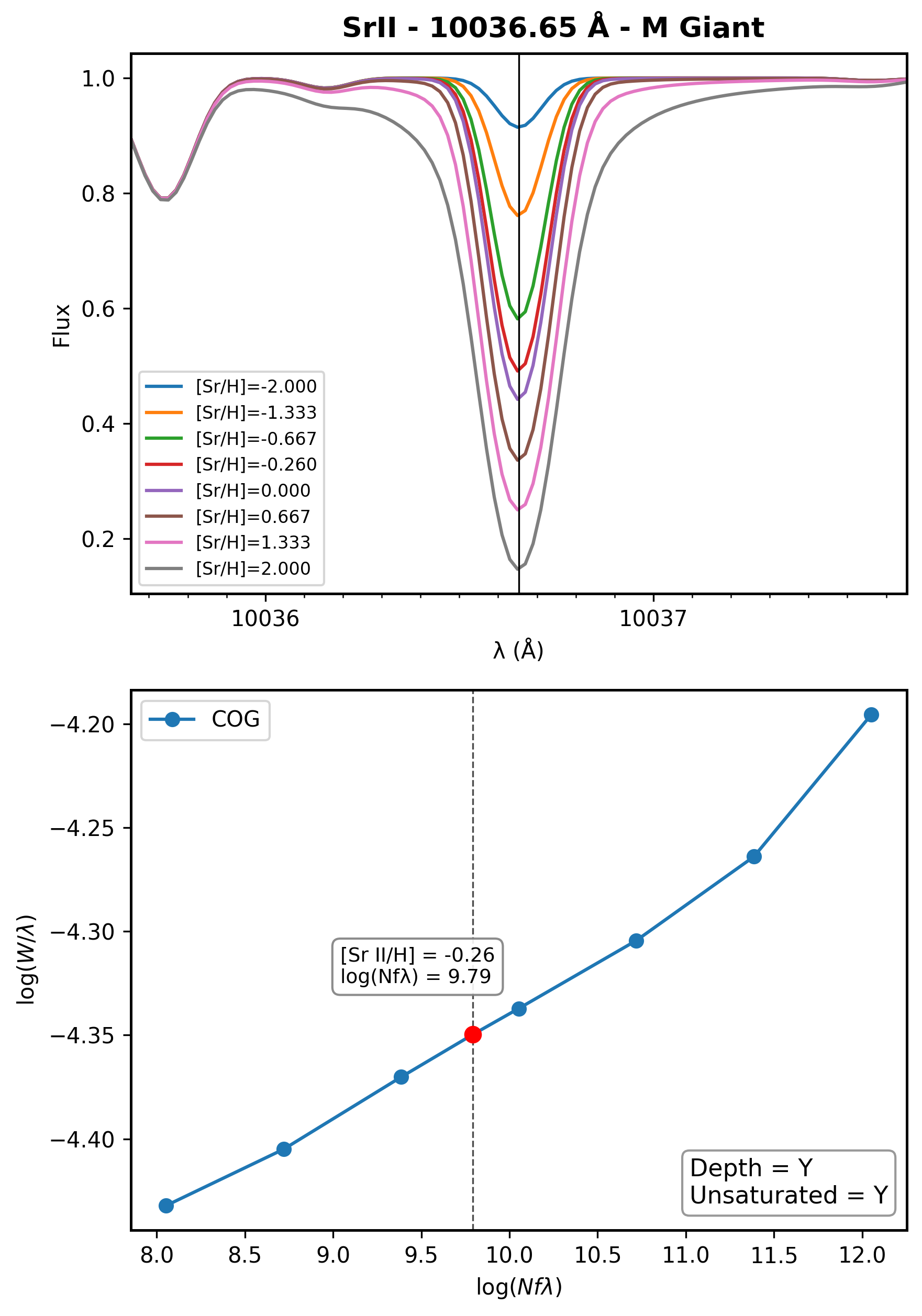}
    \caption{Example of a non-saturated line, the absorption feature of the Sr\,\textsc{ii} line at 10036.65\,\AA\ in a M Giant star. 
\textit{Top panel:} Synthetic spectra computed for different Sr\,\textsc{ii} abundances, ranging from $[{Sr\,\textsc{ii}/H}]=-2.0$ to $+2.0$\,dex. 
\textit{Bottom panel:} Corresponding curve of growth (COG). Each point represents a synthetic spectrum at a given abundance offset. The red dot marks the reference abundance used for the stellar model, i.e., the point where the local COG slope is evaluated to diagnose whether the line is saturated. The dashed vertical line indicates the corresponding $\log(Nf\lambda)$ value. }

    \label{fig:Sr2_MGiant_cog}
\end{figure}

From each synthesis, we measure the equivalent width by numerical integration,
${\rm EW}=\int (1-F_\lambda)\,{\rm d}\lambda$, and form the classical COG variables:
\[
\log\left(\frac{W}{\lambda}\right) \quad\text{vs.}\quad
X \equiv \log(N f \lambda) \simeq \log gf + A + \log\lambda ,
\]
where $f$ is the oscillator strength and $N$ the number density implied proxied by $A$.

\begin{figure*}
    \centering
    \includegraphics[width=\linewidth]{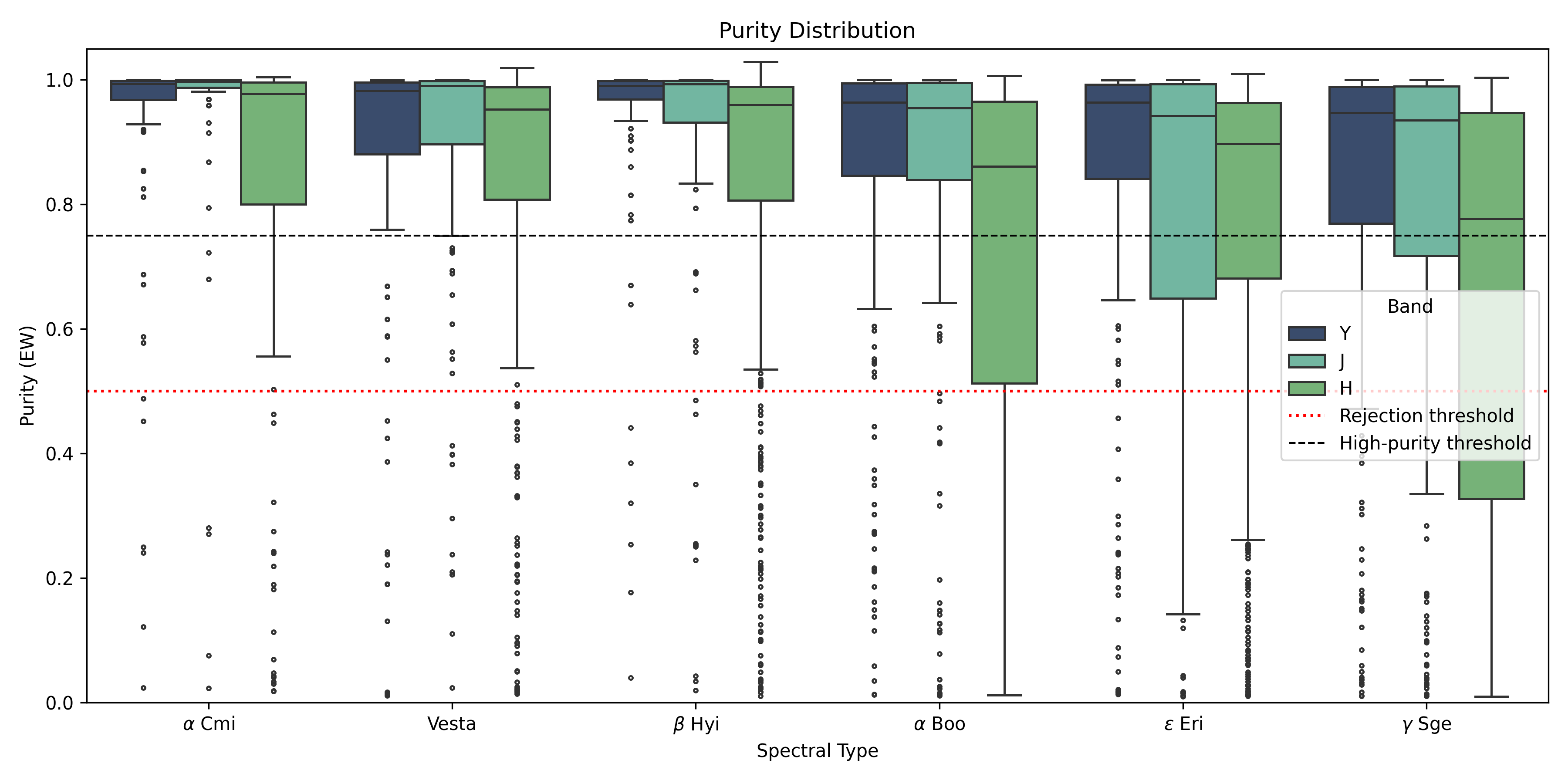}
\caption{
Distribution of purity values (P) for all lines with purity flags N, U and Y. The dashed horizontal line indicates the high-purity threshold at $P = 0.75$, whereas the red dotted line indicates the rejection limit.).
}
    \label{fig:U_Y_purity}
\end{figure*}

We first fit a straight line to all $(X,\log(W/\lambda))$ points to capture the overall behaviour of the curve of growth. 
Then we evaluate a local slope around the reference abundance in optical, to check how the line actually responds at the abundance we adopt for our stars.  This way, we see both the global trend and the sensitivity at the reference point. Figure~\ref{fig:Sr2_MGiant_cog} shows this for the Sr \,\textsc{ii} line: the line core deepens with abundance, and the red point marks the adopted reference abundance ($[{\rm Sr II/H}] = -0.26$\,dex) with its associated $\log(N f \lambda) = 9.79$. The depth and saturation flags indicate that this line meets the depth criterion \texttt{Y}, and is not saturated \texttt{Y}, making it ready for the next step, as shown in Table~\ref{tab:depth_purity_gof_flags}.

In contrast, a line is flagged as Unsaturated \texttt{N} when the local COG slope falls below a threshold
(${\rm d}\log(W/\lambda)/{\rm d}X < 0.5$ in our implementation) and the change in the core flux between
adjacent abundances is negligible ($\Delta F_{\rm core}<0.005$), 
showing the relationship between $\log(W/\lambda)$ and $\log(N f \lambda)$.

The sequential logic, summarized in the flowchart of Figure~\ref{fig:ir_flow}, ensures that only lines with sufficiently strong absorption profiles are carried forward in the line selection process, thereby minimizing the influence of poorly defined or noise-dominated features on the robustness of the final list.

\subsubsection{Purity}
\label{sec:purity}

To evaluate the presence of blended lines within the bands defined in Section~\ref{sec:obs}, we focus on stellar features identified as deeper than 0.03, and not saturated as defined in Sections~\ref{subsub:depth} and~\ref{subsec:saturation}. For these lines, we estimate their degree of blending by measuring two equivalent widths over velocity ranges centered at the line wavelength ($\lambda_c$): $\Delta_1 = 30$~km\,s$^{-1}$ and $\Delta_2 = 60$~km\,s$^{-1}$, resulting in $EW_1$ and $EW_2$, respectively.

Because the integration windows are defined in velocity space, their wavelength extent scales as $\Delta\lambda=\lambda_c\,\Delta v/c$.
Thus, $\Delta_1$ corresponds to a total width of $\simeq 1.0$\,\AA\ at 10000 \,\AA\ and $\simeq 1.5$ \,\AA\ at 15000 \,\AA\ (i.e., $\pm0.5$\,\AA\ and $\pm0.75$\,\AA\ around $\lambda_c$), while $\Delta_2$ corresponds to $\simeq 2.0$\,\AA\ at 10000 \,\AA\ and $\simeq 3.0$\,\AA\ at 15000 \,\AA\ (i.e., $\pm1.0$\,\AA\ and $\pm1.5$\,\AA\ around $\lambda_c$). This method follows the prescription of \citet{Kondo-2019}, in which $EW_1$ traces an inner window around the line center (core, in an
operational sense, i.e., the line center plus immediate inner wings), whereas $EW_2$ provides a wider window that is more sensitive to absorption from nearby contaminant features.
We note that, for very strong lines with extended damping wings or for intrinsically broadened profiles (e.g. rotation or strong macroturbulent broadening), the absorption can extend well beyond $\Delta_1$ (and even
$\Delta_2$), so interpreting $EW_1$ as a strict core window is less appropriate. However, such cases are largely mitigated here because we exclude saturated features. Overall, this two-window approach provides a
direct assessment of line isolation within the synthetic spectrum.

Thus, we define our diagnostic tool for purity based on synthetic spectra generated with the atmospheric parameters listed in Table~\ref{tab:gbs_all_vertical}. Following the prescription of \citet{Kondo-2019}, the purity parameter ($P$) is defined as the ratio between the equivalent widths measured in the inner and outer velocity windows introduced above,
\begin{equation}
P \equiv \frac{EW_1}{EW_2}.
\end{equation}
Since $EW_2$ is measured over a wider velocity interval, it naturally includes the absorption captured in $EW_1$ together with additional contributions from the line wings and the surrounding spectral region. Therefore, even for isolated, non-saturated lines, $P$ approaches but does not reach unity. Lower values of $P$ reflect an increasing sensitivity to absorption from nearby contaminant features within the broader window.

In practice, $P$ is computed for each transition at $\lambda_c$ using the same stellar parameters and line lists employed throughout this work, and a higher $P$ indicates a line less affected by blends.
Following the flagging scheme described in Section~\ref{sec:flagging} and illustrated in Figure~\ref{fig:ir_flow}, lines with $P<0.5$ are rejected
and flagged as \texttt{N}, those with $0.5 \leq P < 0.75$ are considered undecided and flagged as \texttt{U}, and those with $P \geq 0.75$ are accepted and flagged as \texttt{Y}. A $P$ value of 0.75 implies that
contaminating features contribute up to 25\% of the total absorption within $\Delta_1$; if left unmodeled, such a percentage would introduce a systematic error of $\sim$ 0.12 dex in the derived abundances. Nevertheless, this threshold represents a practical compromise given the
high density of molecular features in the IR regime.

\begin{table*}
  \centering
  \footnotesize
  \setlength{\tabcolsep}{5pt}
  \caption{Depth, saturation, purity, and goodness-of-fit (GoF) summary for the Si\,\textsc{i} and Sr\,\textsc{ii} features in the $Y$ band across the 
  stars.}
  \label{tab:depth_purity_gof_flags}
  \begin{tabular}{ccccccccccccc}
    \hline\hline
    \noalign{\vskip 3pt}
    $\lambda$ [\AA] & Star & $d_{syn}$ & $d_{obs}$ & \makecell{Flag\\(Depth)} & \makecell{Flag\\(Unsat.)} & \makecell{Purity\\(P)} & \makecell{Flag\\(P)} & RMSE & MAD  & \makecell{\(\chi^2_\nu\)} & \makecell{\(\chi^2_{\nu,\mathrm{ref}}\)} & \makecell{Flag\\(GoF)} \\
    \noalign{\vskip 3pt}
    \hline
    \rule{0pt}{2.8ex}%
    \multirow{6}{*}{\makecell{Si I \\ 10786.849 }}
      & $\alpha$ Cmi & 0.452 & 0.417  & Y & Y & 0.988 & Y & 0.028 & 0.008 & 1.96 & 1.423 & Y \\
      & Vesta & 0.510  & 0.525     & Y & Y & 0.986 & Y & 0.024 & 0.009 & 1.425 & 1.430 & Y \\
      & $\epsilon$ Eri & 0.551 & 0.491 & Y & Y & 0.982 & Y & 0.045 & 0.024 & 5.055 & 1.425 & N  \\
      & $\beta$ Hyi  & 0.490 & 0.478 & Y & Y & 0.988 & Y & 0.019 & 0.005 & 0.91 & 1.423 & Y \\
      & $\alpha$ Boo & 0.527  & 0.556  & Y & Y & 0.927 & Y & 0.032 & 0.011 & 2.569 & 1.433 & Y \\
      & $\gamma$ Sge & 0.538 & 0.582 & Y & Y & 0.936 & Y & 0.035 & 0.016 & 3.168 & 1.425 & Y \\
      \noalign{\vskip 3pt}
    \hline
    \rule{0pt}{2.8ex}%
    \multirow{6}{*}{\makecell{Sr II \\ 10036.653 }}
      & $\alpha$ Cmi & 0.393 & 0.291  & Y & Y & 0.999 & Y & 0.037 & 0.004 & 3.476 & 1.360 & N \\
      & Vesta & 0.294 & 0.310      & Y & Y & 0.997 & Y & 0.021 & 0.007 & 1.165 & 1.365 & Y  \\
      & $\epsilon$ Eri & 0.420 & 0.311 & Y & Y & 0.988 & Y & 0.048 & 0.007 & 5.899 & 1.363 & N \\
      & $\beta$ Hyi & 0.391 & 0.297   & Y & Y & 0.998 & Y & 0.028 & 0.01 & 1.958 & 1.360 & Y \\
      & $\alpha$ Boo & 0.508 & 0.420 & Y & Y & 0.987 & Y & 0.038 & 0.009 & 3.735 & 1.358 & N \\
      & $\gamma$ Sge & 0.391 & 0.297 & Y & Y & 0.956 & Y & 0.036 & 0.027 & 3.217 & 1.358 & N \\
      \noalign{\vskip 3pt}
    \hline
  \end{tabular}
  \tablefoot{
    Columns: (1) Central wavelength; (2) Star; (3) Synthetic Depth; (4) Observed Depth; (5) Depth flag (Y if both depths $\geq 0.03$); 
    (6) Non-saturation flag (Y if line is not saturated); (7) Purity value (P); 
    (8) Purity flag (Y if $P \ge 0.75$); (9) RMSE; (10) MAD; (11)$\chi^2_\nu\ $ ; (12) $\chi^2_{\nu,\mathrm{ref}}$ ;
    (13) GoF flag.
  }
\end{table*}

The results of the purity evaluation are summarized in Figure~\ref{fig:U_Y_purity}, which shows the distribution of \(P\) values for all lines that passed the depth and saturation criteria (\(P \geq 0.5\)) for all the stars and bands, providing context for the lines that are kept in this stage. The figure does not include the rejected ($P < 0.5$) lines, since the purpose of the plot is to characterize the behavior of the lines that remain eligible for further analysis, thus we provide a clearer view of relevant regimes for each of the spectral types. Table~\ref{tab:depth_purity_gof_flags} provides a concrete example of this multi-step selection process, showing how only those lines with sufficient depth and responsiveness to abundance variations are passed to the purity stage. 

\subsubsection{Goodness of Fit}
\label{sec:gof}

The final step in our line-selection procedure consists in a quantitative evaluation of the agreement between observed and synthetic line profiles at high spectral resolution.
Rather than relying solely on visual inspection, which can be oftentimes subjective, we define a set of statistical metrics computed over a window of $\pm 1.0$\,\AA\ around each line center. Within this window, we evaluate three simple and reproducible statistics: the reduced chi-square ($\chi^2_{\mathrm{red}}$), the root mean square error (RMSE), and the median absolute deviation (MAD). To avoid interpolation artifacts, neither spectrum is resampled. Instead, we pair each observed flux point with the nearest synthetic value (allowing for a minimal wavelength shift, $\delta\lambda$, to account for residual radial velocity offsets). For each match we define the residual as $r_k = f^{\rm obs}_k - f^{\rm syn}_k$.; all reported metrics are computed from these residuals.

\begin{figure*}[h!]
  \centering
  \includegraphics[width=\linewidth]{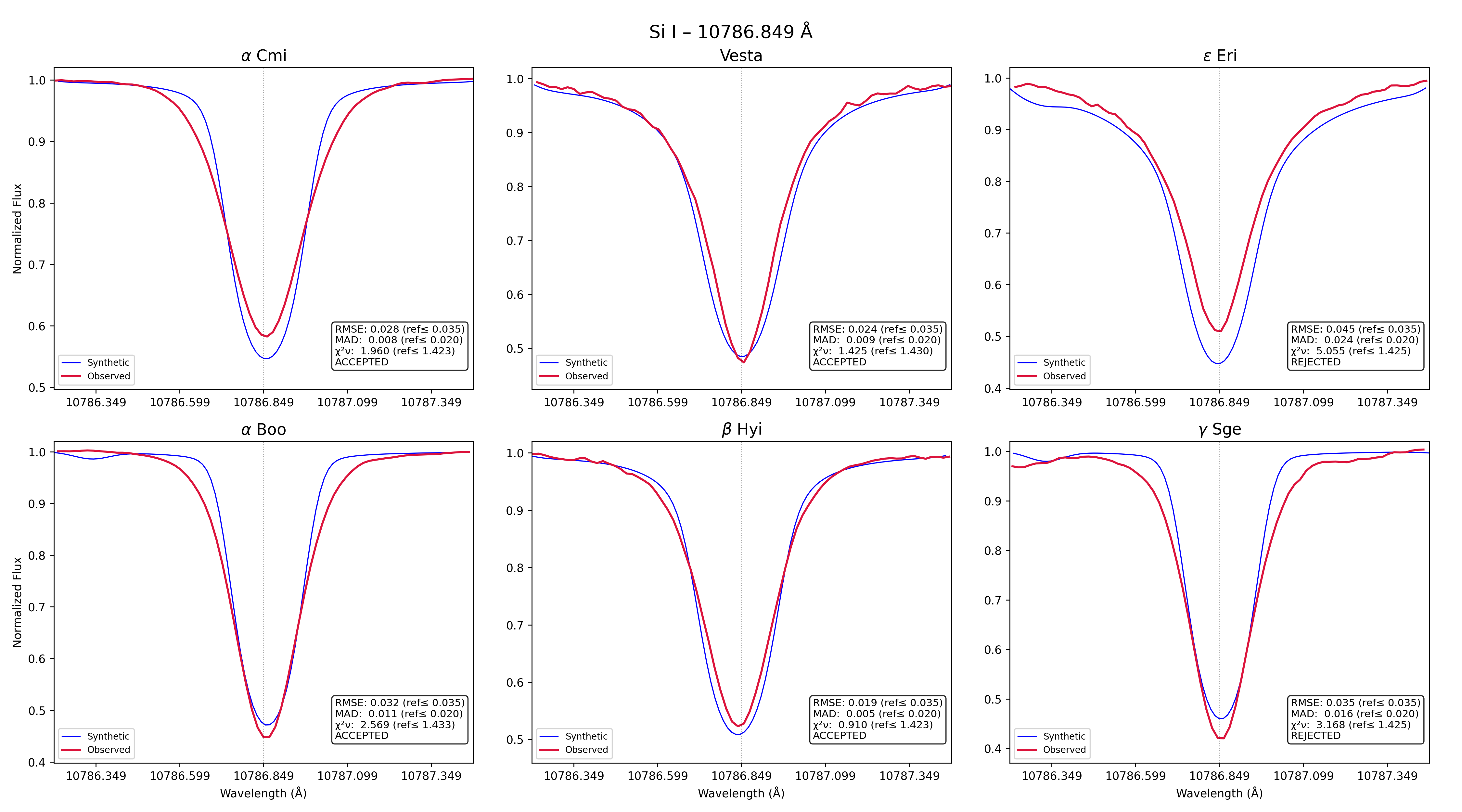}\par\vspace{0.75em}
  \includegraphics[width=\linewidth]{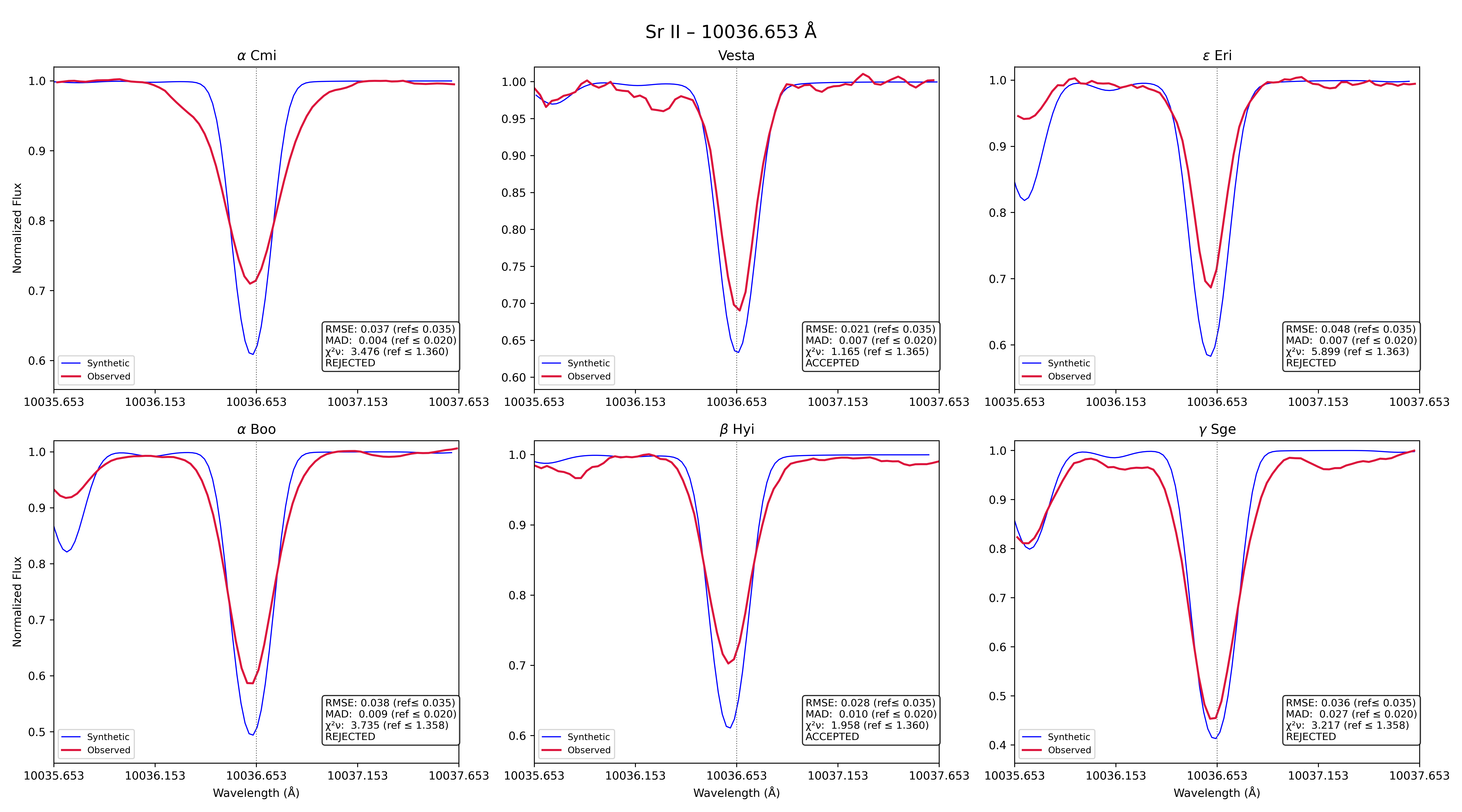}
  \caption{Observed (red) versus synthetic (blue) spectra for two representative infrared lines stacked vertically. Each panel shows six stellar targets and reports the reduced chi-square ($\chi^2_{\mathrm{red}}$), RMSE, and MAD used for the GoF evaluation. The different outcomes for $\alpha$~Boo and $\beta$~Hyi reflect variations in line strength and local blending conditions.}
  \label{fig:gof-stacked-si-sr}
\end{figure*}
A critical component of our methodology is the recognition that imposing a uniform statistical threshold across the entire NIR regime is physically inadequate. The Earth's atmosphere imposes a telluric absorption pattern that varies fundamentally in nature between the bands.

In the Y band ($9800 - 10800\,$ \AA), telluric contamination is primarily driven by molecular Oxygen ($O_2$) and weak water vapor lines. Since $O_2$ is a stable, well-mixed gas, its absorption features are systematic and easier to correct, leaving Gaussian-like residuals. However, the scenario changes drastically in the J and H bands. These windows are heavily affected by water vapor ($H_2O$) and hydroxyl (OH) airglow emission. Unlike oxygen, the concentration of precipitable water vapor is highly variable on short timescales (minutes), creating erratic residuals that are often non-Gaussian and asymmetric.

Consequently, applying the strict statistical criteria suitable for the stable Y band to the H band would result in a high rate of false negatives. We would risk rejecting astrophysically valid atomic lines simply because they overlap with telluric noise that is notoriously difficult to correct perfectly. To address this, we adopt a differential GoF approach: while we maintain strict thresholds for the Y band to ensure high precision, we allow for slightly more tolerant boundaries in the J and H bands, particularly regarding the RMSE, to accommodate the unavoidable residuals of variable telluric absorption.

The degrees of freedom are computed per star and per line as the number of matched wavelength points minus one, $\nu = N - 1$, where $N$ is the total number of matched wavelength pairs within the adopted window. For the $\pm1.0$\,\AA\ window used in this work, $N$ typically ranges between $\sim$20 - 50 points, depending on the local spectral sampling and masking. This quantity enters the computation of the reduced chi-square, while RMSE and MAD are computed directly from the residuals.

From the residuals $\{r_k\}$, we compute the reduced chi-square ($\chi^{2}_{red}$) by normalizing the sum of the squared residuals by the degrees of freedom ($\nu$):

\begin{equation}
\begin{aligned}
\chi^2_{\mathrm{red}} \; &=\; \frac{1}{\nu}\sum_{k=1}^{N}\!\left(\frac{r_k}{\sigma_k}\right)^2, \\
\mathrm{RMSE} \; &=\; \sqrt{\frac{1}{N}\sum_{k=1}^{N} r_k^2}, \\
\mathrm{MAD} \; &=\; \mathrm{med}\!\left(\left|r_k-\mathrm{med}(r_k)\right|\right).
\end{aligned}
\end{equation}

For the Y band, we adopt fixed reference limits $\mathrm{RMSE}_{\rm ref}=0.035$ and $\mathrm{MAD}_{\rm ref}=0.02$ because, at resolving power $R\sim90000$ (instrumental FWHM $\approx \lambda/R$), the continuum scatter of our normalized spectra is typically at the level of $\sim 2\%$, thus $\mathrm{MAD}_{\rm ref}=0.02$ reflects the noise floor, while $\mathrm{RMSE}_{\rm ref}=0.035 \approx (1.5\text{--}2)\sigma$ accommodates the residual effects of sampling, normalization uncertainties, and unresolved broadening, yet still provides enough discriminatory power to exclude fits that are clearly inconsistent with the observed profiles. The $\chi^2$ threshold is set star-by-star through the degrees of freedom $\nu = N - 1$, using the reduced form $\chi^2_{0.95}(\nu)/\nu$; we allow for a minimal $10\%$ relaxation of this bound to avoid excluding cases that are statistically consistent but lie very close to the critical value, the results of this set of statistics are shown in Table~\ref{tab:depth_purity_gof_flags}.

For the J band, the selection criteria were slightly relaxed compared to the Y band to account for the systematic uncertainties introduced by telluric contamination, which is pervasive in this spectral window. Even after telluric correction, non-Gaussian residuals from atmospheric absorption lines (mainly $\mathrm{H_2O}$ and $\mathrm{CH_4}$) artificially inflate the noise statistics. We therefore increased the reference thresholds to $\mathrm{MAD}_{\rm ref}=0.024$ and $\mathrm{RMSE}_{\rm ref}=0.038$. The increase in the MAD threshold reflects an estimated $\sim 20\%$ elevation in the effective noise floor due to telluric residuals compared to the cleaner Y band. Similarly, the $\mathrm{RMSE}_{\rm ref}$ was adjusted to tolerate localized deviations caused by these remnants without compromising the rejection of lines with fundamentally incorrect profiles or global depth mismatches.

For the H band, typical goodness-of-fit values for FGK stars are $\mathrm{MAD}\sim0.03$ and $\mathrm{RMSE}\sim0.04$. Because of the presence of numerous weak molecular features, some line profiles in this band show residual structure that cannot be fully reproduced by the synthetic models. For this reason, the acceptance thresholds were set to $\mathrm{MAD}=0.05$ and $\mathrm{RMSE}=0.05$.

This normalization ensures that the metric is independent of the number of the data points sampled across the line profile. We accept a line as a good fit if at least two of the following three conditions are met, with thresholds adjusted for the specific noise properties of each band:

\begin{enumerate}
    \item The RMSE is at or below the reference noise floor \( \mathrm{RMSE} \le \mathrm{RMSE}_{\mathrm{ref}} \) .
    \item The MAD is at or below its reference value \( \mathrm{MAD} \le \mathrm{MAD}_{\mathrm{ref}} \) .
    \item The reduced chi-square ($\chi_{\nu}^{2}$) does not exceed the 95th-percentile threshold by more than 10\%.
\end{enumerate}

Using a $\pm 1.0$\,\AA\ window on blend-free lines, rather than restricting the comparison to a narrower region around the line core, increases the number of matched wavelength points and leads to a more stable evaluation of $\chi^2_{\mathrm{red}}$. This choice also improves the estimation of the noise from the local continuum and results in a more reliable determination of the optimal wavelength offset, $\delta\lambda^\star$. A profile may still appear imperfect while being statistically accepted, reflecting limitations in the modeling or in the available atomic data (e.g., uncertainties in $\log gf$, unresolved hyperfine or isotopic structure, or magnetic broadening). Our aim is to retain lines that satisfy an objective and reproducible GoF criterion for subsequent abundance analysis. The entire process of this flagging method can be fully appreciated in Table~\ref{tab:depth_purity_gof_flags}, in which the last columns report the RMSE, MAD, the reduced chi-square $\chi^2_{\mathrm{red}}$, and the GoF flag, providing the final decision in the tree.

In Figure~\ref{fig:gof-stacked-si-sr} we present the comparison between synthetic (blue) and observed (red) profiles for the Si\,\textsc{i} 10786.849~\AA\ and  Sr\,\textsc{ii} 10036.653~\AA\ for our  
sample. We note that for the Si\,\textsc{i} line, $\epsilon$~Eri and $\gamma$ Sge do not show a good agreement between their observed and synthetic spectrum, hence the decision is \texttt{N} (rejected) for those two stars, but accepted for the other four stars. In contrast, the behaviour of the $n$-capture line of Sr\,\textsc{ii} seems to be more complex. Its fit quality varies significantly across the sample, reflecting the differences in line strength, blending conditions, and the limitations of the scaled abundances used in the synthesis. An illustrative case is the comparison between $\alpha$~Boo and $\beta$~Hyi: the line is rejected for $\alpha$~Boo, where it is weaker and more affected by neighboring features, but it is accepted for $\beta$~Hyi, where it remains relatively isolated and reaches the depth and purity required for a reliable fit. These contrasting outcomes emphasize the sensitivity of some Sr\,\textsc{ii} transitions to the local spectral environment and to NLTE effects that remain poorly constrained in the near-infrared.

\section{Robust Selection}
\label{sec:robust}

Because the impact of telluric absorption and blending varies across the near-infrared bands, the quality of a given transition must ultimately be judged within the specific band where it is measured. For this reason, the four criteria introduced above (depth, non-saturation, purity, and GoF) are 
evaluated independently in each band.

A line is considered robust only when it satisfies all four criteria in 
the corresponding band, resulting in a final sequence of \texttt{YYYY}. Lines 
that do not reach this configuration will not be included in the robust list. Appendix~\ref{app:Robust_2}
shows the robust list of the transitions that meet this condition in each band for each of the spectral types analyzed in this study.

In practice, this provides a straightforward and reliable framework for users working with near-infrared observations. These lines can therefore be adopted as trustworthy indicators of chemical abundances, ensuring that subsequent analyses rely only on transitions that are well behaved, minimally affected by blending or telluric residuals, and consistently reproduced by the synthetic spectra.

To further place our robust selections in the context of widely used NIR resources, we performed explicit line-by-line cross-matches against three external compilations: the APOGEE atomic linelist \citep{Smith-2021, Shetrone-2015} and two recent solar-type studies \citep{Marfil-2020, Senturk-2024}. We first cross-matched our robust G-dwarf (Sun) list against the APOGEE atomic linelist and identified 33 transitions in common. The overlap is dominated by Fe\,\textsc{i} (12), with additional matches in Si\,\textsc{i} (9), Mg\,\textsc{i} (4), Ni\,\textsc{i} (3), C\,\textsc{i} (3), and single overlaps in S\,\textsc{i}, Cr\,\textsc{i}, and K\,\textsc{i} (one line each). Cross-matching the FGK-giant (Arcturus) robust list yields 24 common transitions, dominated by Fe\,\textsc{i} (20) with additional overlap in Ni\,\textsc{i} (4). Notably, K\,\textsc{i} is retained in our robust selection across all spectral types; while K\,\textsc{i} lines are known to be sensitive to departures from LTE \citep{Osorio-2020}, APOGEE’s DR17 spectral libraries explicitly include an NLTE treatment for K (together with Na, Mg, and Ca), which supports treating K\,\textsc{i} as a carefully modelled diagnostic within the APOGEE framework.

We then cross-matched our robust G-dwarf list with the compilation presented in \citet{Senturk-2024}, which was constructed for solar-type stars. We find 22 transitions in common: Fe\,\textsc{i} (9), Si\,\textsc{i} (9), Ca\,\textsc{i} (2), and Ti\,\textsc{i} (1). 
Finally, we compared our robust FGK-Giant (Arcturus) list with the Fe-only selection of \citet{Marfil-2020}, in which we identify 15 Fe\,\textsc{i} transitions in common with our robust list.

Taken together, these cross-matches show that a non-negligible subset of our robust transitions coincides with independently established NIR selections, while the remaining fraction reflects the additional constraints imposed in this work (depth, saturation, blending/purity, and profile agreement), as well as our emphasis on robustness across stellar parameter space rather than optimization for a single spectral type. 

To provide a compact input-output view of our selection pipeline, Appendix~\ref{app:in_out} summarizes the line-count throughput of the decision tree. In Table~\ref{tab:inout_yjh}, the Total column reports, for each transition and band, the number of transitions available in the initial VALD3 line list within our working wavelength coverage, while the six stellar-class columns report the number of transitions that pass all selection criteria adopted in this work for that class (i.e. the final robust YYYY outcome). For clarity, we omit species with zero accepted transitions in all stellar classes. This table therefore, provides an immediate, quantitative measure of how strongly the selection affects each species as a function of band and spectral type.

\section{Discussion and concluding remarks}
\label{sec:discussion}

A solid evaluation of infrared absorption lines requires combining measurements of the observables with statistical fit metrics. This set of metrics and criteria may not fully capture how a line responds across different stellar parameters. Hence, evaluating the suitability of each line for the GBS sample requires a comprehensive assessment that reflects the underlying physics that shape each transition.

The line selection framework developed in this work provides a robust, reproducible pathway for identifying high-fidelity atomic lines for abundance studies in the near-infrared, leveraging depth, saturation, and purity criteria, as well as a strict, statistical fit-quality assessment across a broad grid of representative stellar types. This approach enables the objective selection of lines based on well-defined and reproducible criteria, minimizing user choices, and maintaining an adaptable nature to meet the needs of mid- to high-resolution infrared spectroscopy.

Nonetheless, one must acknowledge that a residual component of uncertainty persists, particularly related to the complexity of the spectral modeling itself. This modeling uncertainty encompasses a variety of factors, including, but not limited to: the adopted oscillator strengths ($\log gf$) values, the choice and structure of stellar atmosphere models, assumptions regarding LTE versus NLTE conditions, and the role of magnetic fields in line broadening, as demonstrated in \cite{Hahlin-2023}. 

The case of $\gamma$~Sge deserves a dedicated remark. Unlike the other stars in our sample, $\gamma$~Sge is an M giant with an effective temperature of $\sim$3900~K, a regime in which the spectrum becomes substantially more complex. Molecular absorption dominates large portions of the near-infrared, the continuum is difficult to define, and the atmospheric parameters reported in the literature still lack unanimous agreement. As a consequence, even small variations in $T_{\mathrm{eff}}$, $\log g$, or metallicity translate into noticeable differences in the depth and shape of atomic features, complicating the comparison between synthetic and observed profiles. In this sense, $\gamma$~Sge serves as a reminder of the practical limits of any line-assessment methodology: even a rigorous, step-by-step selection framework will inevitably face reduced predictive power when applied to cool giants whose spectra are intrinsically more sensitive to model assumptions and chemical inhomogeneities.

A particularly notable challenge in this work was the inclusion of $n$-capture elements within the robust final list. Although some of them such as Sr\,\textsc{ii} or Ce\,\textsc{ii} are present in the full VALD linelist, their detectability at significant depth and purity is highly restricted to specific spectral types (F and G Dwarf). Under our robustness criterion, most $n$-capture candidates were unfortunately excluded, with Sr\,\textsc{ii} being the one that passed the filters for the Y band for many of the spectral types. This outcome reflects the inherent difficulty in compiling a comprehensive list of clean $n$-capture features for large-scale abundance studies in the infrared. Extending the list of reliable $n$-capture lines may require relaxing the occurrence threshold, increasing the permissible range of line depths, or designing a strategy that yields a better decision when scaling the abundances when they are not measured in the optical. A more flexible strategy, involving a relaxed occurrence threshold or tailored abundance scaling for stars where optical measurements are unavailable, will be explored in a dedicated follow-up study.

Beyond the elements that display a consistently robust behaviour across all spectral types, our analysis also provides insight into species for which the near-infrared regime plays a particularly critical role. This is the case for life-bearing elements such as P\,\textsc{i} and S\,\textsc{i}, whose observational constraints differ substantially despite their similar astrophysical relevance.

While S\,\textsc{i} has been widely investigated across a wide range of stellar types and is relatively well anchored by optical abundance analyses, P\,\textsc{i} remains far less constrained. In particular, P\,\textsc{i} lacks a robust empirical calibration, and only a limited number of transitions are accessible in the near-infrared at high spectral resolution. Moreover, although we initially explored a broader set of near-infrared transitions, our final selection is dominated by $\alpha$ and Fe-peak species, since many other candidates do not satisfy our uniform quality criteria (depth, saturation, purity and gof). For several of these non-standard species, including P\,\textsc{i}, homogeneous optical abundances are not available for the Gaia benchmark stars \citep[e.g.][]{Casamiquela-2025}, so the near-infrared is not merely complementary, but provides the primary route to assess the diagnostic power and internal consistency of their measurable transitions.

It is worth discussing our choice to avoid the widely used APOGEE line list as a primary source. Although APOGEE's line selection has been invaluable for large-scale Galactic surveys, as well as paving the way for the advent of MOONS, its calibration is primarily astrophysical, optimized empirically to match observed spectra rather than tied directly to laboratory data. For our purposes, which demand a standardized, laboratory-based approach for consistent comparison and reproducibility, reliance on astrophysical $\log gf$ calibrations would introduce potential biases across different stellar populations and observing setups. Instead, our methodology emphasizes a uniform and reproducible assessment of line quality, based solely on measurable criteria. Because these criteria depend only on the atomic data and on the observed and synthetic spectra, the selection can be systematically updated whenever improved laboratory measurements, revised atomic parameters, or higher-quality observations become available.

Overall, the cross-matches presented above in Section~\ref{sec:robust} highlight an important point: our goal is not to reproduce the full content of previous line lists, since our selection strategy is different and more conservative. Within that context, the fact that we still recover a clear overlap with widely used references and with independent solar-type studies is reassuring, because it shows that our robust selection converges on many of the same reliable NIR abundance diagnostics identified by other methodologies. 
At the same time, our line list is not limited to those overlaps: it also provides additional robust transitions that are absent from those external compilations, reflecting the wider parameter coverage explored here and the uniform acceptance criteria applied throughout. Together, these aspects support the use of our final recommended lines as a secure reference for homogeneous abundance measurements in the \textit{Y}, \textit{J}, and \textit{H} bands, including applications to stars whose properties are less well studied and constrained.

Looking ahead, there are several avenues for expanding the scope of the robust line list. Lowering the already modest minimum depth threshold could admit additional weak but measurable features, though this must be balanced against increasing vulnerability to blends and noise. Another consideration is the further refinement of $\log gf$ values, either by adopting more recent laboratory determinations or by tailoring values empirically for individual spectral types when robust laboratory values are unavailable. Given the direct degeneracy between oscillator strength and abundance, errors in  $\log gf$ manifest primarily as offsets in the line depth for a fixed abundance. Therefore, refining these values is critical for abundance accuracy, although it has a lesser impact on the assessment of blending or continuum placement.

In summary, the procedures and criteria presented here set a new standard for the rigorous identification of robust atomic lines in the near-infrared, offering a pathway toward homogeneous, reproducible, and physically grounded chemical abundance analyses across diverse stellar populations.

\section{Data Availability}
All the atomic line lists used in this work, including the tables with the transitions analyzed, are available in electronic form.
These tables can be downloaded in machine-readable format alongside the online version of this article.
Additional material will be provided upon reasonable request to the corresponding author.

\section{Acknowledgments}

SE acknowledges financial support from the ANID Millennium Institute of Astrophysics MAS (ICN12\_009), and the CONICYT/ANID FONDECYT Postdoctoral Fellowship No.~3240529.  Additional thanks go to Mingie Jian for providing valuable support with \texttt{PYMOOG}. PJ, CAG and SE acknowledge FONDECYT REGULAR number 1231057. ARA acknowledges support from DICYT through grant 062319RA. MZ acknowledges FONDECYT Regular 1230731 and ANID BASAL Center for
Astrophysics and Associated Technologies (CATA) FB210003. CAG acknowledges FONDECYT Iniciacion 11230741. UH acknowledges support from the Swedish National Space Agency (SNSA/Rymdstyrelsen). SE acknowledges Jerome P. De Leon for the insightful conversations throughout the development of this research.

%
%




\bibliographystyle{aa} 
\bibliography{bib.bib} 

\onecolumn
\begin{appendix} 

\section{Robust lines}
\label{app:Robust_2}

\subsection{F Dwarf}
\begin{table}[h]
\centering
\footnotesize
\renewcommand{\arraystretch}{1.25}
\caption{Robust (\texttt{YYYY}) transitions for F Dwarf in the Y, J, and H bands.}
\label{tab:robust_fdwarf}
\begin{tabular}{l l}
\toprule
\multicolumn{2}{c}{\textbf{Y band (F Dwarf) -- 18 lines}} \\
C\,{\sc i} & 10123.870, 10541.240, 10729.530 \\

Fe\,{\sc i} & 9944.207 \\

N\,{\sc i} & 10112.481, 10114.640 \\

Ni\,{\sc i} & 9898.928 \\

P\,{\sc i} & 10596.903 \\

S\,{\sc i} & 10455.470, 10459.460 \\

Si\,{\sc i} & 10025.743, 10582.160, 10585.141, 10603.425, 10749.378, 10784.562, 10786.849 \\

Sr\,{\sc ii} & 10327.311 \\

\midrule
\multicolumn{2}{c}{\textbf{J band (F Dwarf) -- 16 lines}} \\
C\,{\sc i} & 11801.080, 11819.040, 11863.010, 11879.580, 11892.910, 11895.750, 12569.040, 12614.100 \\

Ca\,{\sc ii} & 11838.997, 11949.744 \\

Si\,{\sc i} & 11863.920, 11984.198, 11991.568, 12103.534, 12270.692, 13152.743 \\

\midrule
\multicolumn{2}{c}{\textbf{H band (F Dwarf) -- 14 lines}} \\
C\,{\sc i} & 16021.700, 16890.380, 17321.610 \\

Cr\,{\sc i} & 15860.210 \\

Fe\,{\sc i} & 17166.201, 17282.291 \\

K\,{\sc i} & 15168.376 \\

Mg\,{\sc i} & 15024.997, 15040.246, 17108.631 \\

Si\,{\sc i} & 15361.161, 15376.831, 15888.409, 15960.063 \\

\bottomrule
\end{tabular}
\end{table}

\clearpage

\subsection{G Dwarf}

\begin{table}[h]
\centering
\footnotesize
\renewcommand{\arraystretch}{1.25}
\caption{Robust (\texttt{YYYY}) transitions for G Dwarf in the Y, J, and H bands.}
\label{tab:robust_gdwarf}
\begin{tabular}{l l}
\toprule
\multicolumn{2}{c}{\textbf{Y band (G Dwarf) -- 46 lines}} \\
C\,{\sc i} & 10123.870, 10541.240, 10683.080, 10685.340, 10691.250, 10707.320, 10729.530, 10753.980 \\

Ca\,{\sc i} & 10343.819 \\

Cr\,{\sc i} & 9949.073, 10510.010, 10647.640, 10667.520, 10672.140 \\

Fe\,{\sc ii} & 10501.503 \\

Fe\,{\sc i} & 10674.070, 10785.387 \\

Ni\,{\sc i} & 9898.928, 10530.535 \\

P\,{\sc i} & 10529.524, 10581.577 \\

S\,{\sc i} & 10456.790, 10459.460 \\

Si\,{\sc i} & 9839.328, 9872.207, 9887.047, 9891.725, 10025.743, 10068.329, 10156.129, 10371.263, 10582.160, 10585.141 \\
 & 10603.425, 10627.648, 10660.973, 10689.716, 10694.251, 10727.406, 10749.378, 10784.562, 10786.849 \\

Sr\,{\sc ii} & 10036.653, 10327.311 \\

Ti\,{\sc i} & 10496.116, 10726.390 \\

\midrule
\multicolumn{2}{c}{\textbf{J band (G Dwarf) -- 42 lines}} \\
C\,{\sc i} & 11801.080, 11848.710, 11863.010, 11879.580, 11892.910, 11895.750, 12549.490, 12562.120, 12581.590, 12601.490 \\
 & 12614.100 \\

Ca\,{\sc ii} & 11949.744 \\

Ca\,{\sc i} & 11955.955, 12823.867 \\

Cr\,{\sc i} & 12532.840 \\

Fe\,{\sc i} & 11970.497, 11989.546, 12283.298, 12667.112, 13014.841, 13147.920 \\

Mg\,{\sc i} & 11828.171 \\

Ni\,{\sc i} & 12216.544 \\

Si\,{\sc i} & 11863.920, 11900.055, 11984.198, 11987.112, 11991.568, 12031.504, 12080.391, 12081.972, 12103.534, 12110.659 \\
 & 12178.339, 12189.241, 12270.692, 12390.154, 12395.832, 12439.963, 13152.743 \\

Ti\,{\sc i} & 12831.445 \\

Zn\,{\sc i} & 13053.627 \\

\midrule
\multicolumn{2}{c}{\textbf{H band (G Dwarf) -- 40 lines}} \\
C\,{\sc i} & 16021.700, 16419.330, 16890.380, 17045.180, 17427.440 \\

Cr\,{\sc i} & 15860.210 \\

Fe\,{\sc i} & 16051.734, 16073.870, 16345.494, 16377.388, 16446.550, 16466.921, 16665.481, 16843.228, 16892.384, 16898.883 \\
 & 16928.623, 16994.742, 17052.181, 17130.952, 17166.201, 17478.016 \\

K\,{\sc i} & 15168.376 \\

Mg\,{\sc i} & 15024.997, 15040.246, 15047.714, 15740.706 \\

Ni\,{\sc i} & 16536.156, 16550.383, 16584.438 \\

S\,{\sc i} & 15400.077 \\

Si\,{\sc i} & 15361.161, 15376.831, 15557.778, 15797.441, 15884.453, 15888.409, 16170.164, 16346.857, 16957.794 \\

\bottomrule
\end{tabular}
\end{table}

\clearpage
\subsection{K Dwarf}
\begin{table}[h]
\centering
\footnotesize
\renewcommand{\arraystretch}{1.25}
\caption{Robust (\texttt{YYYY}) transitions for K Dwarf in the Y, J, and H bands.}
\label{tab:robust_kdwarf}
\begin{tabular}{l l}
\toprule
\multicolumn{2}{c}{\textbf{Y band (K Dwarf) -- 64 lines}} \\
Ca\,{\sc i} & 10273.684, 10343.819, 10558.425 \\

Cr\,{\sc i} & 9900.909, 9946.320, 10080.350 \\

Fe\,{\sc i} & 9811.504, 9834.185, 9868.186, 9886.081, 9889.035, 9924.388, 9977.641, 9980.463, 10019.793, 10026.080 \\
 & 10041.472, 10058.253, 10065.045, 10081.393, 10086.242, 10114.014, 10145.561, 10149.076, 10195.105, 10216.313 \\
 & 10230.795, 10265.217, 10283.775, 10362.704 \\

Ni\,{\sc i} & 9898.928, 10302.611, 10330.248, 10530.535 \\

S\,{\sc i} & 10455.470, 10456.790, 10459.460 \\

Si\,{\sc i} & 9887.047, 10371.263, 10585.141, 10603.425, 10660.973, 10689.716, 10694.251, 10727.406, 10741.728, 10749.378 \\
 & 10784.562, 10786.849 \\

Sr\,{\sc ii} & 10036.653, 10327.311 \\

Ti\,{\sc i} & 9832.140, 9927.350, 9997.959, 10003.088, 10011.744, 10034.492, 10059.905, 10460.050, 10496.116, 10552.965 \\
 & 10661.623, 10677.047, 10732.865 \\

\midrule
\multicolumn{2}{c}{\textbf{J band (K Dwarf) -- 28 lines}} \\
C\,{\sc i} & 12601.490 \\

Ca\,{\sc ii} & 11949.744 \\

Ca\,{\sc i} & 11955.955, 12827.059 \\

Cr\,{\sc i} & 12532.840 \\

Fe\,{\sc i} & 12044.129, 12053.082, 12131.178, 13147.920 \\

Mg\,{\sc i} & 11828.171, 12433.452 \\

Ni\,{\sc i} & 12216.544, 13048.181 \\

Si\,{\sc i} & 11863.920, 11900.055, 11984.198, 11991.568, 12031.504, 12103.534, 12110.659, 12133.995, 12270.692, 12390.154 \\
 & 13152.743 \\

Ti\,{\sc i} & 11892.876, 12671.095, 12811.478, 12831.445 \\

\midrule
\multicolumn{2}{c}{\textbf{H band (K Dwarf) -- 50 lines}} \\
Ca\,{\sc i} & 15067.041 \\

Cr\,{\sc i} & 15680.060, 15860.210, 16015.320 \\

Fe\,{\sc i} & 15522.607, 15551.433, 15560.784, 15566.725, 15571.749, 15590.046, 15591.490, 15631.948, 15652.871, 15670.124 \\
 & 15673.151, 15682.513, 15691.853, 15692.747, 15723.586, 15798.559, 15901.518, 15920.642, 15980.725, 16051.734 \\
 & 16073.870, 16153.247, 16258.912, 16284.769, 16345.494, 16377.388, 16460.368, 16466.921, 16486.666, 16537.994 \\
 & 16928.623, 17130.952, 17166.201, 17221.395 \\

K\,{\sc i} & 15168.376 \\

Mg\,{\sc i} & 15024.997, 15040.246, 15047.714 \\

Ni\,{\sc i} & 15555.375 \\

Si\,{\sc i} & 15376.831, 16163.691, 16957.794 \\

Ti\,{\sc i} & 15602.842, 15715.573, 16635.158, 17376.574 \\

\bottomrule
\end{tabular}
\end{table}

\clearpage

\subsection{FGK Giant}

\begin{table}[h]
\centering
\footnotesize
\renewcommand{\arraystretch}{1.25}
\caption{Robust (\texttt{YYYY}) transitions for FGK Giant in the Y, J, and H bands.}
\label{tab:robust_fgkgiant}
\begin{tabular}{l l}
\toprule
\multicolumn{2}{c}{\textbf{Y band (FGK Giant) -- 40 lines}} \\
C\,{\sc i} & 10123.870, 10685.340, 10707.320, 10729.530 \\

Ca\,{\sc i} & 10273.684, 10558.425 \\

Fe\,{\sc i} & 9811.504, 9820.241, 9886.081, 9944.207, 9953.470, 10041.472, 10227.994, 10230.795, 10333.184 \\

Mg\,{\sc i} & 9828.132 \\

S\,{\sc i} & 10456.790, 10459.460 \\

Si\,{\sc i} & 9872.207, 10288.944, 10313.197, 10371.263 \\

Sr\,{\sc ii} & 10036.653 \\

Ti\,{\sc i} & 9832.140, 9879.582, 9927.350, 9997.959, 10003.088, 10005.661, 10011.744, 10034.492, 10048.826, 10057.728 \\
 & 10059.905, 10120.895, 10170.486, 10551.756, 10552.965, 10565.953, 10677.047 \\

\midrule
\multicolumn{2}{c}{\textbf{J band (FGK Giant) -- 22 lines}} \\
Ca\,{\sc ii} & 11838.997 \\

Ca\,{\sc i} & 11955.955, 12827.059 \\

Fe\,{\sc i} & 11990.386, 12267.888, 12340.481, 12648.741, 12824.859 \\

Mg\,{\sc i} & 12423.029 \\

Si\,{\sc i} & 12103.534, 12270.692, 12390.154, 12395.832, 12627.674, 13154.531, 13176.888 \\

Ti\,{\sc i} & 11949.547, 12167.245, 12255.702, 12811.478, 12919.899, 12987.567 \\

\midrule
\multicolumn{2}{c}{\textbf{H band (FGK Giant) -- 40 lines}} \\
C\,{\sc i} & 17323.470, 17338.560, 17455.990 \\

Cr\,{\sc i} & 14978.730 \\

Fe\,{\sc i} & 15112.331, 15118.115, 15294.560, 15305.603, 15343.788, 15496.697, 15524.308, 15566.725, 15590.046, 15864.646 \\
 & 15901.518, 16284.769, 16312.904, 16345.494, 16377.388, 16537.994, 16721.462, 16837.877, 16892.384, 16928.623 \\
 & 17037.787, 17052.181, 17094.434, 17130.952, 17166.201, 17221.395, 17277.479, 17293.642, 17400.579 \\

Ni\,{\sc i} & 14834.885, 15173.584, 16013.745, 16136.096, 16673.582 \\

Ti\,{\sc i} & 17383.097, 17446.745 \\

\bottomrule
\end{tabular}
\end{table}

\clearpage
\subsection{FGK Subgiant}

\begin{table}[h]
\centering
\footnotesize
\renewcommand{\arraystretch}{1.25}
\caption{Robust (\texttt{YYYY}) transitions for FGK Subgiant in the Y, J, and H bands.}
\label{tab:robust_fgksubgiant}
\begin{tabular}{l l}
\toprule
\multicolumn{2}{c}{\textbf{Y band (FGK Subgiant) -- 43 lines}} \\
Ca\,{\sc ii} & 9854.759 \\

Ca\,{\sc i} & 10273.684, 10343.819 \\

Cr\,{\sc i} & 10486.250 \\

Fe\,{\sc i} & 9834.185, 9944.207, 9951.157, 9977.641, 9980.463, 10041.472, 10114.014, 10362.704, 10423.027, 10435.355 \\
 & 10775.499 \\

Mg\,{\sc i} & 9986.475, 9993.209, 10312.531 \\

Ni\,{\sc i} & 10530.535 \\

P\,{\sc i} & 10084.277, 10529.524, 10596.903 \\

S\,{\sc i} & 10455.470, 10456.790, 10459.460 \\

Si\,{\sc i} & 9887.047, 9969.132, 10025.743, 10156.129, 10371.263, 10414.913, 10582.160, 10585.141, 10603.425, 10660.973 \\
 & 10689.716, 10694.251, 10727.406, 10749.378, 10784.562, 10786.849 \\

Sr\,{\sc ii} & 10327.311 \\

Ti\,{\sc i} & 10726.390 \\

\midrule
\multicolumn{2}{c}{\textbf{J band (FGK Subgiant) -- 25 lines}} \\
C\,{\sc i} & 11801.080, 11819.040, 11863.010, 11879.580, 12569.040, 12581.590, 12601.490, 12614.100 \\

Ca\,{\sc ii} & 11838.997, 11949.744 \\

Ca\,{\sc i} & 11955.955 \\

Fe\,{\sc i} & 13077.337 \\

Mg\,{\sc i} & 11828.171, 12417.937 \\

Si\,{\sc i} & 11863.920, 11900.055, 11984.198, 11991.568, 12080.391, 12081.972, 12103.534, 12270.692, 12390.154, 12583.924 \\
 & 13152.743 \\

\midrule
\multicolumn{2}{c}{\textbf{H band (FGK Subgiant) -- 49 lines}} \\
C\,{\sc i} & 16465.110, 16505.210, 16954.120, 17045.180 \\

Cr\,{\sc i} & 15680.060 \\

Fe\,{\sc i} & 15301.557, 15305.603, 15360.234, 15387.803, 15496.697, 15514.279, 15551.433, 15566.725, 15571.749, 15652.871 \\
 & 15901.518, 15982.072, 16051.734, 16073.870, 16171.930, 16258.912, 16284.769, 16345.494, 16377.388, 16384.141 \\
 & 16466.921, 16476.933, 16720.727, 16724.685, 16863.431, 16928.623, 17052.181 \\

K\,{\sc i} & 15168.376 \\

Mg\,{\sc i} & 15040.246 \\

Ni\,{\sc i} & 15173.584, 15199.616, 16013.745, 16550.383 \\

P\,{\sc i} & 16482.932 \\

Si\,{\sc i} & 15361.161, 15376.831, 15557.778, 15797.441, 15827.213, 15888.409, 16163.691, 16571.433, 16957.794 \\

Ti\,{\sc i} & 15715.573 \\

\bottomrule
\end{tabular}
\end{table}
\clearpage

\subsection{M Giant}
\begin{table}[h]
\centering
\footnotesize
\renewcommand{\arraystretch}{1.25}
\caption{Robust (\texttt{YYYY}) transitions for M Giant in the Y, J, and H bands.}
\label{tab:robust_mgiant}
\begin{tabular}{l l}
\toprule
\multicolumn{2}{c}{\textbf{Y band (M Giant) -- 39 lines}} \\
Fe\,{\sc i} & 9800.798, 9811.504, 9820.241, 9834.185, 9868.186, 9886.081, 9889.035, 9937.090, 9944.207, 9951.157 \\
 & 9953.470, 9977.641, 9980.463, 9987.868, 10041.472, 10145.561, 10149.076, 10230.795, 10283.775, 10364.062 \\
 & 10455.403, 10535.709, 10611.686, 10674.070, 10682.393, 10735.519, 10754.281, 10771.228 \\

Ni\,{\sc i} & 10302.611 \\

Sr\,{\sc ii} & 10036.653 \\

Ti\,{\sc i} & 9879.582, 9941.378, 9981.275, 10034.492, 10048.826, 10050.244, 10066.512, 10170.486, 10496.116 \\

\midrule
\multicolumn{2}{c}{\textbf{J band (M Giant) -- 30 lines}} \\
Cr\,{\sc i} & 12000.970, 12979.450, 13192.910 \\

Fe\,{\sc i} & 11989.546, 11990.386, 12005.397, 12131.178, 12267.888, 12321.203, 12340.481 \\

Mg\,{\sc i} & 11828.171, 12039.822, 12417.937 \\

Si\,{\sc i} & 11984.198, 11991.568, 12031.504, 12081.972, 12270.692, 12390.154, 12583.924, 13102.058, 13154.531, 13176.888 \\

Ti\,{\sc i} & 12071.707, 12167.245, 12264.257, 12321.872, 12388.373, 13002.195, 13112.673 \\

\midrule
\multicolumn{2}{c}{\textbf{H band (M Giant) -- 18 lines}} \\
Fe\,{\sc i} & 16008.075, 16039.853, 16040.654, 16072.242, 16231.646, 16316.320, 16486.666, 16753.065 \\

Mg\,{\sc i} & 15024.997, 15040.246, 15740.706, 16163.699 \\

Si\,{\sc i} & 16060.008, 16094.787, 16163.691, 16241.833, 16380.176 \\

V\,{\sc i} & 15774.060 \\

\bottomrule
\end{tabular}
\end{table}

\clearpage
\section{Line list input–output summary}
\label{app:in_out}
Table~\ref{tab:inout_yjh} reports the number of transitions per atomic species in the Y, J, and H bands, separated by stellar class. The Total column gives the total number of transitions for each species present in the initial line list, while each stellar-class column reports the number of transitions that pass all selection criteria adopted in this work for that class.

\begin{longtable}{lccccccc}
\caption{In/out line counts per species in the Y, J, and H bands.}\label{tab:inout_yjh}\\
\toprule
Specie & Total & F Dwarf & G Dwarf & K Dwarf & FGK Giant & FGK Subgiant & M Giant \\
\midrule
\endfirsthead

\toprule
Specie & Total & F Dwarf & G Dwarf & K Dwarf & FGK Giant & FGK Subgiant & M Giant \\
\midrule
\endhead

\midrule
\multicolumn{8}{r}{Continued on next page}\\
\midrule
\endfoot

\bottomrule
\endlastfoot
\multicolumn{8}{c}{\textbf{Y band}}\\
\midrule
C\,{\sc i}   & 67   & 3  & 8  & 0  & 4  & 0  & 0  \\
N\,{\sc i}   & 92   & 2  & 0  & 0  & 0  & 0  & 0  \\
Mg\,{\sc i}  & 9    & 0  & 0  & 0  & 1  & 3  & 0  \\
Si\,{\sc i}  & 117  & 7  & 19 & 12 & 4  & 16 & 0  \\
P\,{\sc i}   & 37   & 1  & 2  & 0  & 0  & 3  & 0  \\
S\,{\sc i}   & 183  & 2  & 2  & 3  & 2  & 3  & 0  \\
Ca\,{\sc i}  & 480  & 0  & 1  & 3  & 2  & 2  & 0  \\
Ca\,{\sc ii} & 58   & 0  & 0  & 0  & 0  & 1  & 0  \\
Ti\,{\sc i}  & 1189 & 0  & 2  & 13 & 17 & 1  & 9  \\
Cr\,{\sc i}  & 1281 & 0  & 5  & 3  & 0  & 1  & 0  \\
Fe\,{\sc i}  & 3403 & 1  & 2  & 24 & 9  & 11 & 28 \\
Fe\,{\sc ii} & 4176 & 0  & 1  & 0  & 0  & 0  & 0  \\
Ni\,{\sc i}  & 404  & 1  & 3  & 2  & 2  & 2  & 1  \\
Y\,{\sc ii}  & 205  & 0  & 0  & 0  & 0  & 0  & 0  \\
Sr\,{\sc ii} & 17   & 1  & 1  & 1  & 1  & 1  & 1  \\
\midrule
\multicolumn{8}{c}{\textbf{J band}}\\
\midrule
C\,{\sc i}   & 87   & 8  & 11 & 1  & 0  & 8  & 0  \\
Mg\,{\sc i}  & 31   & 0  & 1  & 2  & 1  & 2  & 3  \\
Si\,{\sc i}  & 173  & 6  & 11 & 6  & 10 & 6  & 8  \\
Ca\,{\sc i}  & 373  & 0  & 1  & 1  & 5  & 1  & 5  \\
Ca\,{\sc ii} & 63   & 2  & 0  & 0  & 1  & 1  & 0  \\
Ti\,{\sc i}  & 1032 & 0  & 0  & 4  & 4  & 0  & 5  \\
Cr\,{\sc i}  & 1149 & 0  & 2  & 2  & 0  & 0  & 0  \\
Fe\,{\sc i}  & 4003 & 0  & 3  & 7  & 1  & 5  & 4  \\
Ni\,{\sc i}  & 530  & 0  & 2  & 0  & 0  & 2  & 2  \\
Zn\,{\sc i}  & 5    & 0  & 0  & 3  & 0  & 0  & 0  \\
\midrule
\multicolumn{8}{c}{\textbf{H band}}\\
\midrule
C\,{\sc i}   & 100  & 3  & 5  & 0  & 3  & 4  & 0  \\
Mg\,{\sc i}  & 156  & 3  & 4  & 3  & 0  & 1  & 4  \\
Si\,{\sc i}  & 408  & 6  & 16 & 11 & 6  & 10 & 1  \\
P\,{\sc i}   & 45   & 0  & 0  & 0  & 0  & 0  & 0  \\
S\,{\sc i}   & 308  & 0  & 1  & 0  & 1  & 0  & 2  \\
K\,{\sc i}   & 60   & 1  & 0  & 2  & 0  & 3  & 0  \\
Ca\,{\sc i}  & 283  & 0  & 2  & 1  & 1  & 1  & 0  \\
Ti\,{\sc i}  & 2157 & 0  & 0  & 0  & 1  & 0  & 0  \\
Cr\,{\sc i}  & 1212 & 1  & 1  & 5  & 0  & 1  & 0  \\
Fe\,{\sc i}  & 5151 & 2  & 4  & 18 & 28 & 24 & 0  \\
Ni\,{\sc i}  & 601  & 0  & 2  & 10 & 0  & 5  & 0  \\

\end{longtable}

\end{appendix}

\end{document}